\documentclass[aps, twocolumn, preprintnumbers, amsmath, amssymb, prl, superscriptaddress]{revtex4-1}

\usepackage[dvipsnames]{xcolor}
\usepackage{graphicx}
\usepackage{bbm}

\newcommand{\BFA}{BaFe$_2$As$_2$}

\newcommand{\be}{\begin{eqnarray}}
\newcommand{\ee}{\end{eqnarray}}

\newcommand{\nn}{\nonumber } 
\newcommand{\Eqref}[1]{Eq.~\eqref{#1}}

\usepackage{color}
\usepackage[colorlinks=true,plainpages=false,linkcolor=blue,urlcolor=blue,citecolor=blue,pdfpagemode=UseNone,pdfstartview=FitBH]{hyperref}

\begin{document}

\title{Spin Waves in Detwinned BaFe$_2$As$_2$}

\author{Xingye Lu}
\email{luxy@bnu.edu.cn}
\affiliation{Center for Advanced Quantum Studies and Department of Physics, Beijing Normal University, Beijing 100875, China}

\author{Daniel D. Scherer}
\affiliation{Niels Bohr Institute, University of Copenhagen, Juliane Maries Vej 30, DK-2100 Copenhagen, Denmark }

\author{David W. Tam}
\affiliation{Department of Physics and Astronomy \& Rice Center for Quantum Materials,
Rice University, Houston, Texas 77005, USA}

\author{Wenliang Zhang}
\affiliation{Beijing National Laboratory for Condensed Matter Physics, Institute of Physics, Chinese Academy of Sciences, Beijing 100190, China}

\author{Rui Zhang}
\affiliation{Department of Physics and Astronomy \& Rice Center for Quantum Materials,
Rice University, Houston, Texas 77005, USA}

\author{Huiqian Luo}
\affiliation{Beijing National Laboratory for Condensed Matter Physics, Institute of Physics, Chinese Academy of Sciences, Beijing 100190, China}

\author{Leland W. Harriger}
\affiliation{NIST Center for Neutron Research, National Institute of Standards and Technology, Gaithersburg, Maryland 20899, USA}

\author{H. C. Walker}
\affiliation{ISIS Facility, Rutherford Appleton Laboratory, Chilton, Didcot, Oxfordshire OX11 0QX, UK}
\author{D. T. Adroja}
\affiliation{ISIS Facility, Rutherford Appleton Laboratory, Chilton, Didcot, Oxfordshire OX11 0QX, UK}
\affiliation{Highly Correlated Matter Research Group, Physics Department, University of Johannesburg, P.O. Box 524, Auckland Park 2006, South Africa}

\author{Brian M. Andersen}
\email{bma@nbi.ku.dk}
\affiliation{Niels Bohr Institute, University of Copenhagen,  Juliane Maries Vej 30, DK-2100 Copenhagen, Denmark }

\author{Pengcheng Dai}
\email{pdai@rice.edu}
\affiliation{Department of Physics and Astronomy \& Rice Center for Quantum Materials,
Rice University, Houston, Texas 77005, USA}
\affiliation{Center for Advanced Quantum Studies and Department of Physics, Beijing Normal University, Beijing 100875, China}

\date{\today}

\begin{abstract}
Understanding magnetic interactions in the parent compounds of high-temperature superconductors forms the basis for determining their role for the mechanism of superconductivity. For parent compounds of iron pnictide superconductors such as $A$Fe$_2$As$_2$ ($A=$ Ba, Ca, Sr), although spin excitations have been mapped out throughout the entire Brillouin zone (BZ), measurements were carried out on twinned samples and did not allow for a conclusive determination of the spin dynamics. Here we use inelastic neutron scattering to completely map out spin excitations of $\sim$100\% detwinned BaFe$_2$As$_2$. By comparing observed spectra with theoretical calculations, we conclude that the spin excitations can be well described by an itinerant model with important contributions from electronic correlations. 
\end{abstract}

\maketitle

It is well-known that high-temperature superconductivity in copper oxides and iron pinctides arises from electron and hole-doping of their antiferromagnetically order parent compounds \cite{PALee06,Keimer,scalapinormp,dairmp}. Since magnetism is believed to be important for superconductivity of these materials \cite{PALee06,Keimer,scalapinormp,dairmp,taconnphys,wangncom}, it is therefore crucial to determine the magnetic interactions in the parent compounds in order to understand their evolution as a function of electron/hole-doping. For insulating antiferromagnetic (AF) copper oxides such as La$_2$CuO$_4$, spin waves can be well described by a local-moment Heisenberg Hamiltonian \cite{headings,Coldea}. In the case of metallic iron pnictide such 
as $A$Fe$_2$As$_2$ ($A=$ Ba, Ca, Sr), a parent of iron-based superconductors, the material exhibits a tetragonal-to-orthorhombic structural transition at $T_s$ and forms twin-domains before ordering antiferromagnetically at $T_N$ ($T_s\ge T_N$) \cite{qhuang,mgkim}. Although spin waves throughout the Brillouin zone (BZ) have been mapped out on twinned samples, 
they do not allow a conclusive determination of the intrinsic magnetic exchange interactions and the origin of 
magnetism due to complications arising from formation of twin-domains, which mixes spin-waves from the twin domains at the same position 
in reciprocal space \cite{dairmp,zhaojun,ames,leland11,Ewings11}.

\begin{figure}
\includegraphics[width=6.6cm]{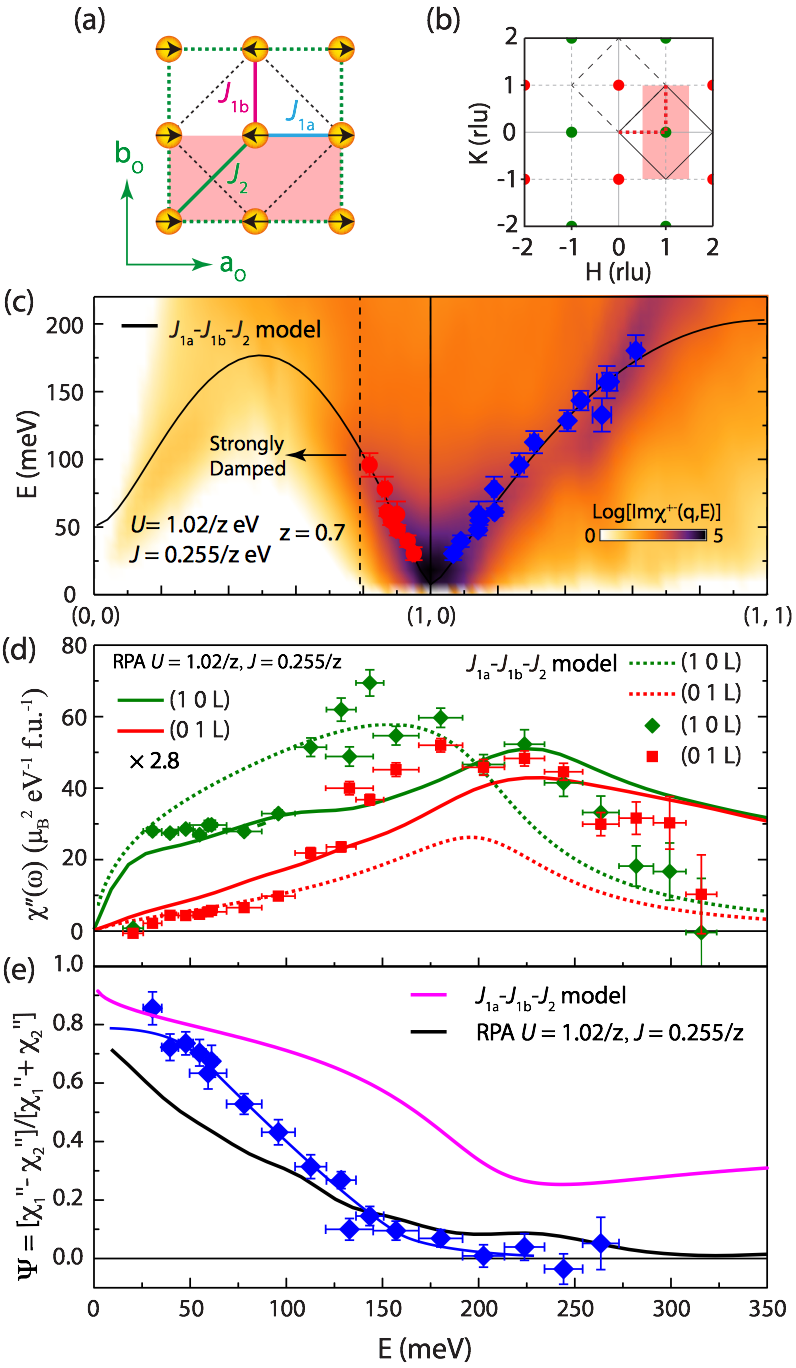}
\caption{Summary of neutron-scattering results and theoretical calculations of spin waves of detwinned {\BFA}.
(a) Spin arrangement of Fe$^{2+}$ in the FeAs plane and the definition of the effective magnetic exchange 
couplings $J_{1a}, J_{1b}$ and $J_2$. The pink area marks the AF unit cell of {\BFA}. (b) Reciprocal space of {\BFA} with twin domains. The green and red dots mark the magnetic Bragg peak positions for twin domains. The pink rectangular area is the AF Brillouin zone. The black diamonds centered at $(1, 0)$ and $(0, 1)$ are the integration region for calculating energy-dependent local dynamic susceptibility. (c) Spin-wave dispersions of a detwinned {\BFA} extracted from constant-energy cuts collected at $T=7$K.  
The black curves are obtained from a Heisenberg model ($J_{1a}-J_{1b}-J_2$) fit of twinned {\BFA} \cite{leland11}. The background shows the spectral weight from the RPA calculation (renormalized with $ z = 0.7 $) with $ U = 1.02 / z $\,eV and $ J = U/4 $ \cite{SI}. (d) Energy-dependent local susceptibility $\chi''(E)$ for AF Brillouin zones at $(1, 0)$ and $(0, 1)$. The green and red dashed lines are spin-wave fits from a Heisenberg Hamiltonian obtained from a twinned sample (with arbitrary unit) \cite{leland11}. The green and red solid lines are from MF+RPA calculations, which were multiplied by 2.8 for clear comparison. (e) Spin-wave anisotropy $\psi(E)$. The purple and black solid lines are calculated spin-wave anisotropy from Heisenberg Hamiltonian and MF+RPA, respectively. The blue line is a guide to the eye of the experimental data. The vertical error bars in (c) and horizontal error bars in (d) and (e) mark the integrating energy ranges. The vertical error bar in (c) indicates $1\sigma$ confidential interval for the fitting of the momentum position. The ones in (d) and (e) originate from the uncertainty of the scattered neutrons and the propagation of the uncertainty for the calculation of $\chi''(E)$ and $\psi$, respectively.}
\label{fig1}
\end{figure}

In this Letter, we report inelastic neutron scattering measurements of spin excitations in uniaxial-strain detwinned BaFe$_2$As$_2$ \cite{fisher11,Dhital12,Lu14,Song15}.  In the unstrained state, {\BFA} undergoes a nearly simultaneous structural and magnetic phase transition at $T_s\approx T_N\approx 138$ K from a paramagnetic tetragonal state to an AF orthorhombic state \cite{qhuang,mgkim}. Below $T_N$, {\BFA} exhibits a collinear AF order [Fig. 1(a)], with an in-plane magnetic wave vector $\mathbf{Q}_{AF}=(1, 0)$ [Fig. 1(b)] \cite{qhuang}. Because of the twinning effect, magnetic Bragg peaks appear at both $\mathbf{Q}_{AF} = (\pm1, 0)$ and  $(0, \pm1)$. Therefore, spin waves on twinned {\BFA}  stem from both the $\mathbf{Q}_{AF} = (\pm1, 0)$ and $(0, \pm1)$ positions, and are four-fold symmetric \cite{zhaojun, ames, leland11, Ewings11}. Although spin waves from twinned samples were described by a local-moment Heisenberg Hamiltonian with effective exchange couplings $J_{1a}$, $J_{1b}$, and $J_2$ [Fig. 1(a)]  \cite{zhaojun, leland11}, one can hardly justify the assumption that magnetic excitations will be absent at $(0, \pm1)$ up to the (magnetic) band top in this itinerant system. On the other hand, spin waves from the twin domains overlap at energies close to the band top and therefore make it difficult to determine if a local-moment Heisenberg Hamiltonian can faithfully describe the intrinsic spin-wave spectra of a detwinned sample.

\begin{figure*}[htbp!]
\includegraphics[width=15 cm]{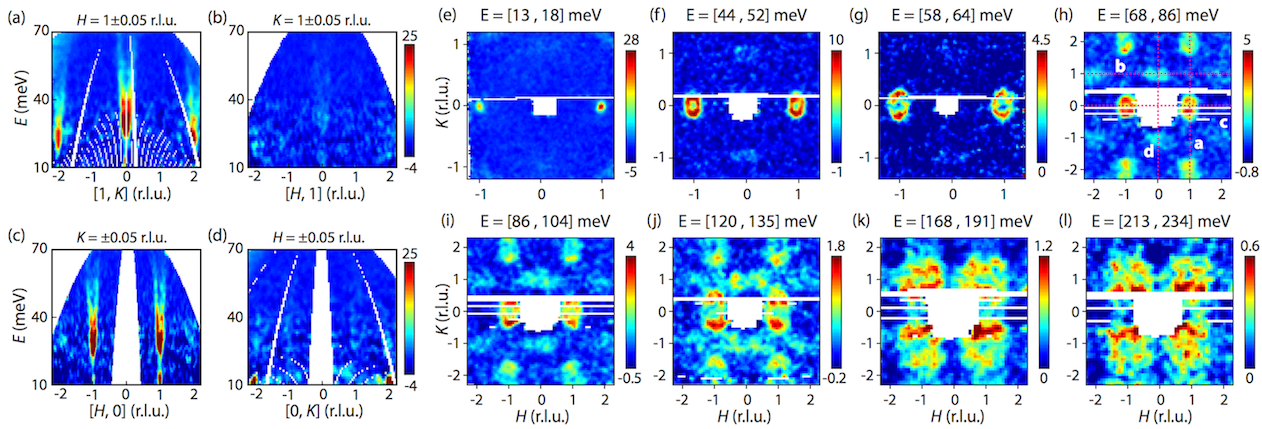}
\caption{Projection of the magnetic scattering intensity ($\frac{d^2\sigma}{d{\mathbf{\Omega}}dE}\frac{k_i}{k_f}$) onto energy and momentum planes. (a)-(d) Magnon dispersions along (a) $[1, K]$, (b) $[H, 1]$, (c) $[H, 0]$ and (d) $[0, K]$ directions measured with $E_i=81$ meV. These four directions are marked by red dashed lines in (h). (e)-(l) Constant energy slices in $[H, K]$ plane. (e) is measured with $E_i=30$ meV, (f)-(g) with $E_i=81$ meV, (h)-(j) with $E_i=250$ meV and (k)-(l) with $E_i=450$ meV. All the data in this figure were collected at $T=7$ K.}
\end{figure*}

To resolve this problem and completely determine the intrinsic spin-wave spectra of detwinned {\BFA}, we carried out inelastic neutron scattering measurements on an assembly of mechanically detwinned {\BFA} single crystals, with pressure ranging from $12$-$22$ MPa \cite{SI}. Our measurements were carried out at MERLIN time-of-flight neutron-scattering spectrometer at ISIS Facility, Rutherford Appleton Laboratory. The sample set was aligned with the $c$-axis along the incident 
beam (${\bf k_i}\parallel c$) direction.

\begin{figure*}[htbp!]
\includegraphics[width=15cm]{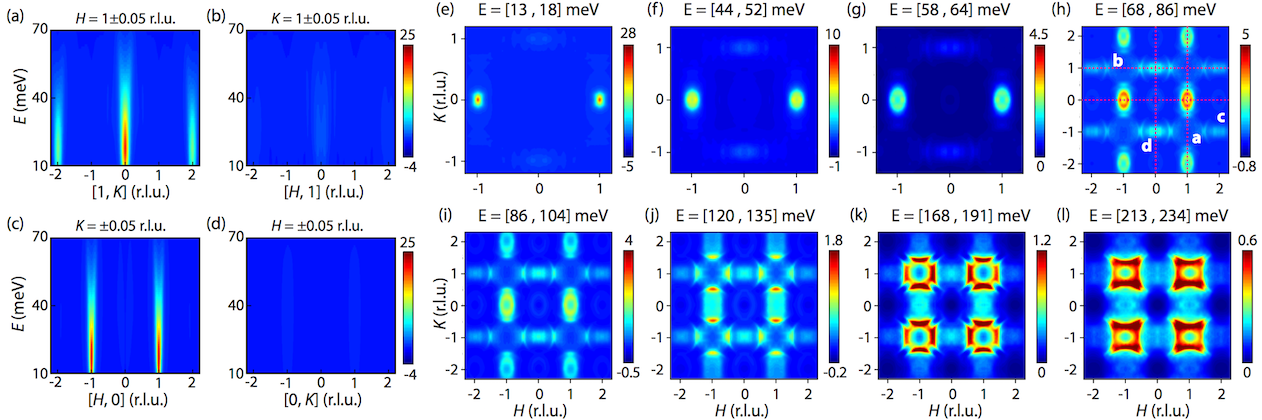}
\caption{Theoretical MF+RPA calculations of the magnetic scattering intensity as shown in Fig. 2 with $U=1.02/z$ eV and Hund's coupling $J=0.255/z$ eV ($z=0.7$). The intensity for magnetic scattering in this figure is obtained from $\chi''(\mathbf{q}, \omega)$ calculated using MF+RPA, taking into account magnetic form factor, Bose factor etc. \cite{SI}. To facilitate the comparison, the intensity from calculation was multiplied by a factor of $2.8$.}
\end{figure*}

Figures 1(c)-1(e) summarize the key results obtained from our measurements of the spin waves. In a completely detwinned sample, the magnetic unit cell in real space and its corresponding BZ in reciprocal space are plotted as pink regions in Figs.~1(a) and 1(b), respectively.  Low-energy spin waves from the collinear AF order in Fig. 1(a) should stem from $(\pm 1,K)$ with $K=0,\pm 2$ positions in reciprocal space [Fig. 1(b)] \cite{Lu14}. The red and blue data points in Fig. 1(c) show spin-wave dispersions from detwinned {\BFA} along the $[H, 0]$ and $[1, K]$ directions, respectively. The black solid lines are dispersion curves along the same two directions from the $J_{1a}$-$J_{1b}$-$J_2$ Heisenberg Hamiltonian describing spin-wave dispersions of twinned {\BFA} \cite{leland11}. We see that the dispersion for detwinned {\BFA} agrees well with results from the Heisenberg fit to the twinned sample, confirming that the uniaxial pressure used to detwin the samples does not affect the magnetic interactions \cite{Lu16PRB}.

\begin{figure}[htbp!]
\includegraphics[width=7cm]{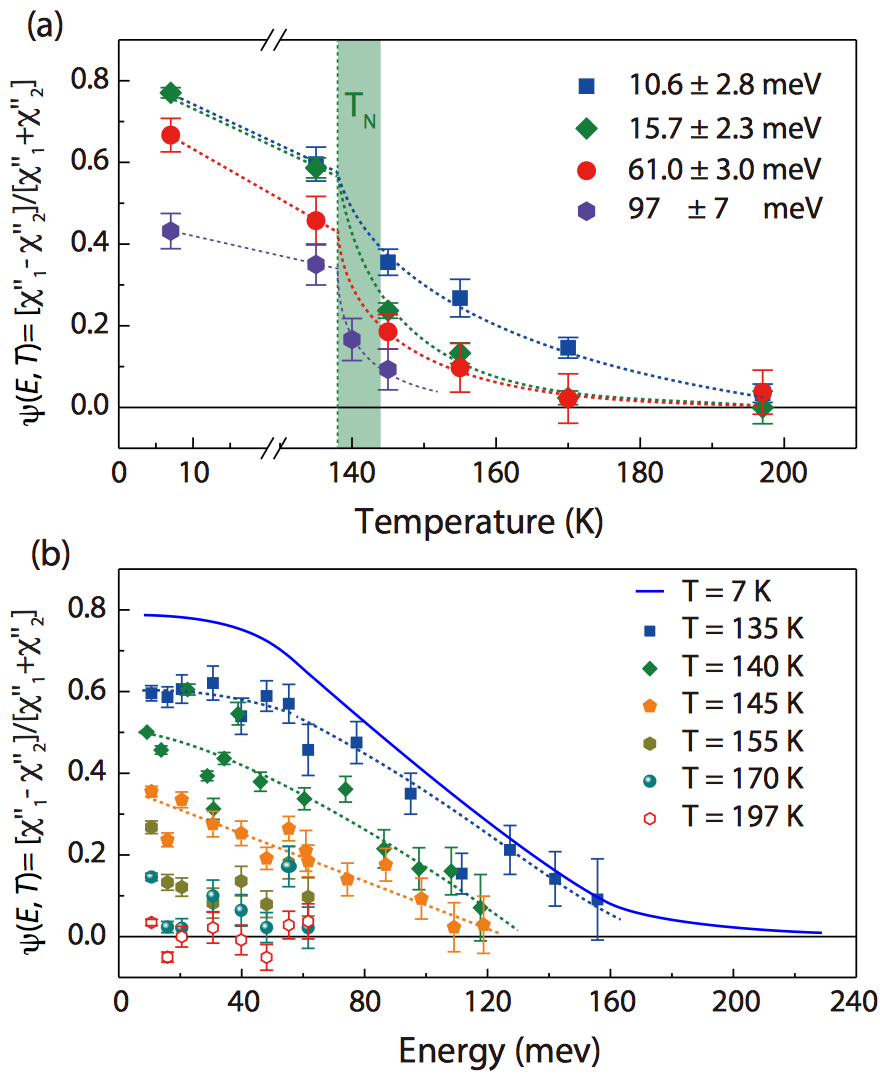}
\caption{ Temperature and energy dependence of the nematic spin correlations of uniaxial-strained {\BFA}. (a) Temperature dependence of the spin-excitation anisotropy ($\psi$) between $(1, 0)$ and $(0, 1)$ for energy transfers of $10.6\pm2.8$, $15.7\pm2.3$, $61\pm3$ and $97\pm7$ meV. (b) $\psi$ as a function of energy transfer measured at various temperatures from $7$ K to $197$ K. The solid line marks the anisotropy for $7$K as shown in Fig. 1(e). The dashed lines in (a) and (b) are guides to the eye. }
\end{figure}

To further test if the Heisenberg Hamiltonian \cite{leland11} can also describe the spin excitations of detwinned {\BFA}, we consider the energy dependence of the local dynamic susceptibility, defined as 
$\chi''(E)=\int_{\rm BZ}\chi^{\prime\prime}({\bf Q},E)d{\bf Q}/\int_{\rm BZ}d{\bf Q}$, where $\chi^{\prime\prime}({\bf Q},E)$ is wave vector and energy dependence of the imaginary part of the dynamic susceptibility within a BZ [pink rectangle or black diamond in Fig. 1(b)] \cite{dairmp}, at $(1,0)$ (denoted by $\chi''_1$) and $(0,1)$ (denoted by $\chi''_2$) wave vectors.  For twinned {\BFA}, $\chi^{\prime\prime}_{1}(E)$ equals to $\chi''_2(E)$ at all energies and spin waves exhibit $C_4$ rotational symmetry \cite{leland11}. The green diamonds and red squares in Fig. 1(d) show the measured $\chi''_1(E)$ and $\chi''_2(E)$ in a detwinned {\BFA}, respectively. In the present study, $\chi''(\mathbf{Q}, E)$ is calibrated using a standard vanadium sample. The averaged $\chi''(E)=[\chi''_1(E)+\chi''_2(E)]/2$ shows the same energy dependence as that for twinned sample \cite{leland11}, with approximately the same intensity (within the $\sim 30\%$ error of the absolute intensity
calibration) \cite{SI}. While the local dynamic susceptibility is dominated by $\chi''_1(E)$ for spin-wave energies below $\sim$100 meV, $\chi''_1(E)$ and $\chi''_2(E)$ become indistinguishable for energies above 170 meV. For comparison, the dashed green and red lines are the corresponding calculated local dynamic susceptibilities using parameters obtained
from fits to spin waves in a twinned sample, which have different $\chi''_1(E)$ and $\chi''_2(E)$ for all energies \cite{leland11}. We see that the Heisenberg Hamiltonian fails at all energies to describe $\chi''_1(E)$ and $\chi''_2(E)$ in a detwinned {\BFA}. In Fig. 1(e), this is shown more clearly in the energy dependence of magnetic susceptibility anisotropy, defined as $\psi(E)=[\chi''_1(E)-\chi''_2(E)]/[\chi''_1(E)+\chi''_2(E)]$. The anisotropy calculated from the Heisenberg Hamiltonian is much larger than experimental results at most energies, because spin excitations arise only from $(1, 0)$ in this picture.

Figures 2(a)-2(i) reveal the energy and wave-vector dependence of the spin excitations in detwinned {\BFA} measured at $T = 7$ K. Figures 2(a) and 2(c) plot the background subtracted spin-wave scattering for $E_i=81$ meV projected in (${\bf Q}, E$) planes with ${\bf Q}$ along the $[1,K]$ and $[H,0]$ directions, respectively. Sharp spin waves are seen to originate from the AF ordering wave vectors $(1,K)$ with $K=0,\pm 2$ [Fig. 2(a)] and $(\pm 1,0)$ [Fig. 2(c)].  Similar projections around wave vectors $(0,\pm 1)$ yield no visible magnetic scattering at the expected twin-domain positions, confirming the nearly 100\% detwinning ratio of the {\BFA} sample, as seen from Figs. 2(b) and 2(d). Figures 2(e)-2(l) show the two-dimensional (2D) constant-energy images of the spin excitations in the $(H,K)$ plane at different energies. For spin-wave energies of $E=15.5\pm 2.5$ meV [Fig. 2(e)], $48\pm 4$ meV [Fig. 2(f)], $61\pm 3$ meV [Fig. 2(g)], we see clear spin-wave rings stemming from ${\bf Q}_{AF}=(\pm 1,0)$ with essentially no observable scattering from the twin-domain positions $(0,\pm 1)$. For spin-wave energies at $E=77\pm 9$ meV [Fig. 2(h)], $97\pm 9$ meV [Fig. 2(i)], $127.5\pm 7.5$ meV [Fig. 2(j)], the spin modes split along the $[1,K]$ direction, and weak spin excitations appear at the $(0,\pm 1)$ positions. Upon further increase of the energy to $E=179.5\pm 11.5$ [Fig. 1(k)] and $223.5\pm 10.5$ meV [Fig. 2(l)], we can no longer identify any spin-wave anisotropy, and the excitations exhibit $C_4$ rotational symmetry as in a twinned sample.

To understand the data presented in Figs. 1 and 2, we model the electronic degrees of freedom with $3d$-Fe orbital character by a multiorbital Hubbard model. The hopping matrix-elements describing the propagation of uncorrelated electrons are taken from the five-orbital model in Ref.~\cite{Ikeda2010}, while the interaction Hamiltonian consists of intra- and inter-orbital onsite repulsion as well as Hund's coupling and a pair-hopping interaction \cite{SI}. Within the framework of a multiorbital Hubbard model, information about the magnetic fluctuations of the electronic system, as probed by inelastic neutron scattering, can be extracted from the electronic spin-spin correlation functions. Here, we determine these correlation functions within the random phase approximation (RPA) that treats the electronic system as composed of coherent quasiparticles, and neglects self-energy effects beyond the mean-field level in general, and the incoherent (and potentially localized) electronic background in particular. The RPA correctly captures the stripe spin-density wave (SDW) instability in the magnetic channel driven by Fermi surface nesting between the electron and hole pockets \cite{Hirschfeld}. The RPA also incorporates Landau-damping effects of the magnetic excitations due to the inclusion of decay into
particle-hole pairs \cite{Kovacic2015}. To account for correlation effects beyond mean-field (MF) theory, we include a phenomenological self-energy that describes both uniform band-renormalization and reduced quasiparticle-weight. The value of the renormalization factor $z$ is then determined by matching the bandwidth of magnetic excitations to the experimental result. The MF+RPA data shown in Figs. 1 and 3 has been renormalized with $ z = 0.7$. This value seems roughly consistent with orbitally averaged estimates from dynamical mean-field theory (DMFT)~\cite{Skornyakov2009} ($ z \approx 0.49 $) and slave-spin mean-field theory~\cite{Medici2014} ($ z \approx 0.43 $) calculations for \BFA.

To correctly capture the Goldstone mode in the magnetic channel when entering the AF phase (where we neglect the spin-rotation-symmetry-breaking effects of spin-orbit coupling, that manifest only at energies less than $\sim$30 meV \cite{NQureshi12, CWangPRX, Yuli}), we self-consistently stabilize a magnetic stripe configuration \cite{Gastiasoro2015, Scherer2016} with a local moment parallel to $ a $ within MF theory and by the RPA determine the magnetic fluctuations in the AF ordered state \cite{Brydon2009b,Knolle2010,Kaneshita2010,Kovacic2015,Scherer2016}. In presenting our results, we limit ourselves to the transverse (with respect to a spin-quantization axis chosen parallel to $ a $) susceptibility. The longitudinal contributions give rise to small quantitative correction only ~\cite{SI}.

Figures 3(a)-3(l) show images of our RPA results at identical energy and wave-vectors as that of the experiments 
in Figs. 2(a)-2(l). The calculated results capture the emergence of the spin excitations at $(0, \pm1)$ and are in reasonable agreement with magnon dispersion and the global topology of the spectral weight distribution, as shown in Figs. 1(c) and 3 \cite{SI}. Figures 2(a) and 2(c) show an intensity maximum at $\sim 30$ meV because of the $L$ modulation of the magnetic excitations, which were not included in the calculation for Fig. 3. Within the RPA, the description of the spin-wave anisotropy improves significantly over the Heisenberg result, as seen from Fig. 1(e). This consistency signals the importance of an itinerant description of the magnetic degrees of freedom in iron pnictides. While at low-energies, Landau-damped spin waves at $(\pm 1, 0 )$ dominate and render the spin excitation spectrum $C_{2}$ symmetric, the spin waves evolve into particle-hole-like excitations for higher energies. The presence of these transverse excitations at both $(\pm1, 0)$ and $(0, \pm1)$ eventually renders the high-energy part of the spectrum $C_{4}$ symmetric and gives rise to a characteristic drop in the spin-wave anisotropy, that cannot be described by the Heisenberg model. 

While the standard RPA-approach is known to yield a too small spectral weight (that translates to a too small fluctuating moment) compared to experiments, we achieve qualitative agreement for the shape of the magnetic excitation spectrum by employing a phenomenological renormalization factor $z$, as can be seen in Fig. 1(c). There is, however, evidence from work on another correlated itinerant system~\cite{gore2018} that the inclusion of vertex corrections is necessary to accurately describe the overall intensity, while the fine-structure of the frequency- and momentum-dependent susceptibility is determined  by the particle-hole propagator.

In addition spin waves, the spin-excitation anisotropy $\psi(E, T)$ above $T_N$ in uniaxial-strained {\BFA} \cite{Lu14}, which is intimately connected to the electronic nematic \cite{chu10, chu12, Ming2011, Ming2017} and reflects the coupling between nematic susceptibility \cite{Lu16PRB} and spin dynamics \cite{Lu14}, has so far only been studied at very low energies \cite{Lu14}. Here, we provide results for the energy and temperature dependence of the spin-excitation anisotropy in the paramagnetic state, which is crucial for understanding the nature of the electronic nematic phase \cite{Fernandes14}.

Figures 4 summarize the temperature and energy dependence of $\psi(E, T)$. Figure 4(a) shows temperature dependence of the spin-excitation anisotropy $\psi(E, T)$ at energies of $E=10.6\pm 2.8$, $15.7\pm 2.3$, $61\pm 3$, and $97\pm 7$ meV. With increasing energy,  $\psi(E, T)$ disappears at progressively lower temperatures, and essentially vanishes above $T_N$ at $E=97\pm 7$ meV. Figure 4(b) shows the energy dependence of $\psi(E, T)$ at temperatures below and above $T_N$. At temperatures 7 K and 135 K ($<T_N$), the spin waves are anisotropic up to $E \approx 160$ meV. Upon warming up to 140 K, 145 K, 155 K, 170 K, and 197 K, the energies of spin-excitation anisotropy decrease gradually with increasing energy and become isotropic at 197 K. These results set an upper limit for the characteristic temperature for the nematic spin correlations, as well as the energy scale of the spin excitations affected by the structural nematic susceptibility \cite{Lu16PRB}.
 
In the paramagnetic phase, MF+RPA calculation gives qualitatively similar results as DFT+dynamical mean-field theory (DMFT) calculations \cite{wangncom,Park2011,Liu2012,Yin2014}, where correlation effects are taken into account on a microscopic level. Since it is challenging to calculate spin waves in the AF ordered state of iron pnictides using DFT+DMFT, the MF+RPA approach allows us to explore the evolution of the spin waves to the paramagnon-like excitations across the AF transition. It turns out, however, that the RPA calculation in the paramagnetic state significantly underestimates the temperature and energy scale of the nematic spin correlations. We attribute the failure of the paramagnetic RPA calculation to capture the observed spin-excitation anisotropy $\psi$ to neglecting the feedback of the temperature-dependent nematic order parameter onto both the electronic states and the spin excitations. Correspondingly, the nematic order parameter obtains a finite value even above $T_{s}$ and therefore can affect both electronic and magnetic properties. Within a spin-nematic scenario \cite{Christensen2016}, the paramagnon-gap at, e.g., $ (\pm 1,0) $ will decrease, while it will increase at $ (0,\pm 1) $. The nematic order will thereby increase the spin-excitation anisotropy compared to our paramagnetic RPA calculation and provide a characteristic temperature dependence.

We thank Changle Liu, Rong Yu, Qimiao Si, and Shiliang Li for helpful discussions.
Part of the work at BNU is supported by  ``the Fundamental Research Funds for the Central Universities'' (X.L.). The neutron-scattering
work at Rice University was supported by the US NSF Grant
No. DMR-1700081 (P.D.). The BaFe$_2$As$_2$ single-crystal
synthesis work at Rice University was supported by the Robert
A. Welch Foundation Grant No. C-1839 (P.D.). D.~D.~S. and B.~M.~A. acknowledge financial support from the Carlsberg Foundation.

%-------------------------------------------------------------------------------------

%------------------------------------------------------------------------------------

\setcounter{equation}{0}
\setcounter{figure}{0}
\setcounter{table}{0}
\setcounter{page}{1}
\makeatletter
\renewcommand{\theequation}{S\arabic{equation}}
\renewcommand{\thefigure}{S\arabic{figure}}

\widetext
\newpage

{~~~~~~~~~~~~~~~~~~~~~\bf SUPPLEMENTAL MATERIAL: Spin Waves in Detwinned {\BFA}}

\maketitle

%%%%%%%%%% Merge with supplemental materials %%%%%%%%%%
%%%%%%%%%% Prefix a "S" to all equations, figures, tables and reset the counter %%%%%%%%%%
\section{MF + RPA Calculation}
%-------------------------------------------------------------------------------------
\subsection{Multiorbital Hubbard model}
\label{sec:model}
%-------------------------------------------------------------------------------------

%
\begin{figure*}[h!]
%-----------------------------------------------------------
\begin{minipage}{0.23\textwidth}
\centering
\flushleft{(a)}
\includegraphics[width=1\columnwidth]{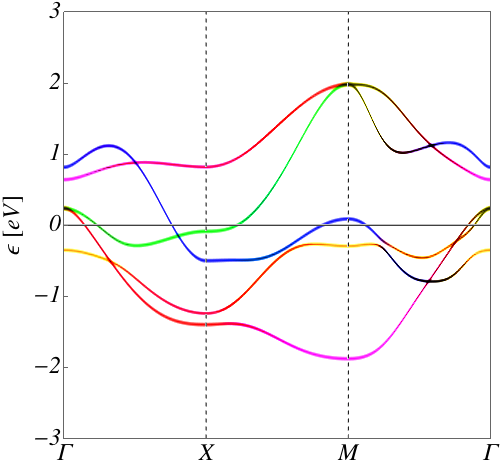}
\end{minipage}
%-----------------------------------------------------------
\begin{minipage}{0.26\textwidth}
\centering
\flushleft{(b)}
\includegraphics[width=1\columnwidth]{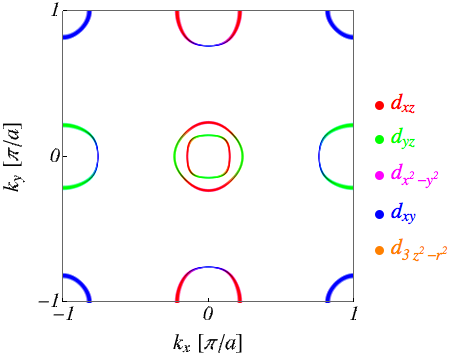}
\end{minipage}
%-----------------------------------------------------------
\begin{minipage}{0.23\textwidth}
\centering
\flushleft{(c)}
\includegraphics[width=1\columnwidth]{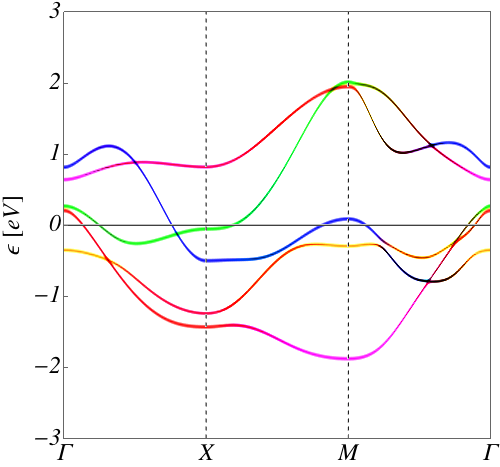}
\end{minipage}
%-----------------------------------------------------------
\begin{minipage}{0.26\textwidth}
\centering
\flushleft{(d)}
\includegraphics[width=1\columnwidth]{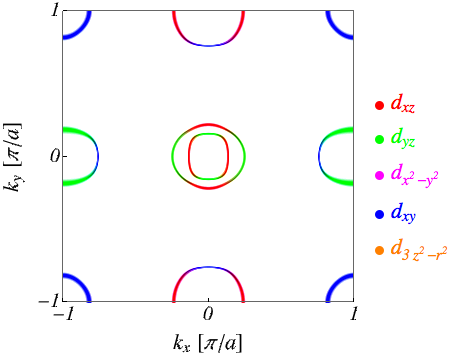}
\end{minipage}
%-----------------------------------------------------------
\caption{(a),(b) Bandstructure and Fermi surface of the 5-orbital model in the absence of orbital splitting. (c),(d) Bandstructure and Fermi surface in the presence of a finite orbital splitting $ \delta = 0.065 $\,eV at the same chemical potential as in (a),(b).}
\label{fig:fs}
\end{figure*}

We consider a $5$-orbital Hubbard model for FeSCs with uniaxial strain. The full Hamiltonian reads
\be
H = H_{0} + H_{\delta} + H_{\mathrm{int}},
\ee
with
\be
\label{eq:hopping}
H_{0} = \sum_{\sigma}\sum_{i,j}\sum_{\mu,\nu} c_{i \mu \sigma}^{\dagger}\left( t_{ij}^{\mu\nu}  - \mu_{0} \delta_{ij}\delta_{\mu\nu} \right)c_{j \nu \sigma},
\ee
describing the hopping of electrons and
\be 
\label{eq:orbitalsplitting}
H_{\delta} = \frac{\delta}{2}\sum_{i}\left( n_{iyz} - n_{ixz} \right),
\ee
a local orbital splitting in the $xz,yz$ manifold to model the effect of uniaxial strain on the electronic spectrum.
The interaction of the electrons is described by a local Hubbard-Hund Hamiltonian,
\be
\label{eq:interaction}
H_{\mathrm{int}} & = &  
U \sum_{i,\mu} n_{i \mu \uparrow} n_{i \mu \downarrow} + 
\left(U^{\prime} - \frac{J}{2}\right) \sum_{i,\mu < \nu, \sigma,\sigma^{\prime}} n_{i \mu \sigma} n_{i \nu \sigma^{\prime}} \nn \\
& & \hspace{-2.5em} 
- 2 J \sum_{i, \mu < \nu}{\bf S}_{i\mu}\cdot{\bf S}_{i\nu}  + 
\frac{J^{\prime}}{2} \sum_{i, \mu \neq \nu,\sigma} \left(c_{i\mu\sigma}^{\dagger}c_{i\mu\bar{\sigma}}^{\dagger}c_{i\nu\bar{\sigma}}c_{i\nu\sigma} + \mathrm{h.c.}\right).
\ee
The hopping elements $t_{ij}^{\mu\nu}$ are taken from \cite{SIikeda2010}. Here we let $\mu,\nu \in \{d_{xz}, d_{yz}, d_{x^2-y^2}, d_{xy}, d_{3z^{2}-r^{2}}\}$ to specify the $3d$-Fe orbitals and $i,j$ run over the sites of the square lattice. The filling is fixed by the chemical potential $\mu_{0}$, and the onsite interaction is parametrized by an intraorbital Hubbard-$U$, an interorbital coupling $U^{\prime}$, Hund's coupling $J$ and pair hopping $J^{\prime}$. We will restrict ourselves to rotationally symmetric interaction parameters, which are realized for $U^{\prime} = U - 2J$, $J = J^{\prime}$. The fermionic operators $ c_{i \mu \sigma}^{\dagger}$, $c_{i \mu \sigma}$ create and destroy, respectively, an electron at site $i$ in orbital $\mu$ with spin polarization $\sigma$. Accordingly, we define the operators for local charge and spin as $n_{i\mu} = n_{i\mu\uparrow} + n_{i\mu\downarrow}$ with $n_{i\mu\sigma} = c_{i\mu\sigma}^{\dagger} c_{i\mu\sigma}$ and ${\bf S}_{i\mu} = 1/2\sum_{\sigma\sigma^{\prime}} c_{i\mu\sigma}^{\dagger} {\boldsymbol \sigma}_{\sigma\sigma^{\prime}}c_{i\mu\sigma^{\prime}}$, respectively. The resulting bandstructure and Fermi surface with and without orbital splitting is shown in Fig.~\ref{fig:fs}.

%-------------------------------------------------------------------------------------
\subsection{Mean-field theory for stripe-SDW state}
\label{sec:meanfield}
%-------------------------------------------------------------------------------------

Following \cite{SIgastiasoro2015,SIscherer2016} we perform a momentum-space mean-field decoupling of the interaction $H_{\mathrm{int}}$. To describe collinear magnetic order, we here restrict ourselves to the mean-fields
\be 
\label{eq:mf1}
n_{0}^{\mu\nu} & = & \frac{1}{\mathcal{N}} \sum_{{\bf k},\sigma} \langle c_{{\bf k}\mu \sigma}^{\dagger}c_{{\bf k}\nu \sigma} \rangle \\
\label{eq:mf2}
M_{1}^{\mu\nu} & = & \frac{1}{\mathcal{N}} \sum_{{\bf k},\sigma} \sigma \langle c_{{\bf k} + {\bf Q_{1}}\mu \sigma}^{\dagger}c_{{\bf k}\nu \sigma} \rangle,
\ee
and neglect spin-off diagonal mean-fields, i.e., $\langle c_{{\bf k} \mu \sigma}^{\dagger} c_{{\bf k}^{\prime} \nu \sigma^{\prime}} \rangle = 0$ for $\sigma \neq \sigma^{\prime}$. We note that we chose a spin-quantization axis $ z = a $, such that the ordered magnetic moment is aligned with the $a$ axis (Fe-Fe direction) of the crystal. The prefactor $\mathcal{N}$ denotes the number of unit cells. The fields $n_{0}^{\mu\nu}$ describe a band renormalization and orbital hybridization, while $M_1^{\mu\nu}$
describes the magnetic order with ordering vector ${\bf Q}_{1} = (\pi,0)$, corresponding to antiferromagnetic arrangement of spins in the $a$-direction, and ferromagnetic alignment along the $b$ direction. Both $n_{0}^{\mu\nu}$ and $M_1^{\mu\nu}$ as well as the chemical potential are determined self-consistently, where the average $\langle \cdots \rangle$ on the right hand sides of \Eqref{eq:mf1} and \Eqref{eq:mf2} is computed from a thermal state of the mean-field Hamiltonian $H_{\mathrm{MF}} = \sum_{{\bf k},\mu,\nu\sigma}^{\prime}\Psi_{{\bf k}\mu}^{\dagger}h^{\mu\nu}({\bf k})\Psi_ {{\bf k}\nu}$. The Bloch-Hamiltonian $h^{\mu\nu}({\bf k})$ with ${\bf k}$ in the reduced Brillouin zone (rBZ) reads as
\begin{eqnarray}
h^{\mu\nu}({\bf k}) = 
\begin{pmatrix}
 \xi^{\mu\nu}({\bf k})  +N^{\mu\nu}_0 & W^{\mu\nu}_1 & 0 & 0 & \\
 W^{\mu\nu}_1 & \xi^{\mu\nu}({\bf k}+{\bf Q}_1)+N^{\mu\nu}_0  & 0 & 0  \\
 0 & 0 & \xi^{\mu\nu}({\bf k})+N^{\mu\nu}_0  & -W^{\mu\nu}_1 \\
 0 & 0 & -W^{\mu\nu}_1 & \xi^{\mu\nu}({\bf k}+{\bf Q}_1) +N^{\mu\nu}_0
\end{pmatrix}, \nn
\end{eqnarray}
where the basis is defined by the spinor
\begin{eqnarray}
\Psi_{{\bf k}\nu}^{\dagger}=
\begin{pmatrix}
c_{{\bf k}\mu\uparrow}^{\dagger} &
c_{{\bf k} + {\bf Q}_{1}\mu\uparrow}^{\dagger} &
c_{{\bf k}\mu\downarrow}^{\dagger} &
c_{{\bf k} + {\bf Q}_{1}\mu\downarrow}^{\dagger}
 \end{pmatrix},
\end{eqnarray}
and $\Psi_{{\bf k}\nu} = (\Psi_{{\bf k}\nu}^{\dagger})^{\dagger}$ The orbital matrices $N_{0}^{\mu\nu}$ and $W_{1}^{\mu\nu}$ entering $h^{\mu\nu}({\bf k})$ are composed of the charge and magnetic mean-fields in Eqs. (\ref{eq:mf1})-(\ref{eq:mf2}) through
\be
\label{eq:N0}
N_{0}^{\mu\nu} & = & 
\delta^{\mu\nu}\Bigl(U n_{0}^{\mu} + (2 U^{\prime} - J) \bar{n}_{0}^{\nu}\Bigr) + 
\bar{\delta}^{\mu\nu} 
\Bigl((-U^{\prime} + 2 J) n_{0}^{\nu\mu} + J^{\prime} n_{0}^{\mu\nu} \Bigr), \\
\label{eq:W1}
W_{1}^{\mu\nu} & = & \delta^{\mu\nu}\Bigl(-U M_{1}^{\mu} - J \bar{M}_{1}^{\nu} \Bigr) + \bar{\delta}^{\mu\nu}
\Bigl( U^{\prime} M_{1}^{\nu\mu} - J^{\prime} M_{1}^{\mu\nu} \Bigr).
\ee
We have introduced further auxiliary quantities to ease the notation, where $\delta^{\mu\nu}$ denotes the Kronecker symbol with respect to orbital indices and $\bar{\delta}^{\mu\nu} = 1 - \delta^{\mu\nu}$ filters out the orbital off-diagonal components. We note, that repeated indices are \emph{not} summed over. Quantities in Eqs.~(\ref{eq:N0})-(\ref{eq:W1}) with a single orbital index refer to the diagonal element of the corresponding matrix, e.g., $ n_{0}^{\mu} = n_{0}^{\mu\mu} $. Objects with a bar, such as $\bar{n}_{0}^{\nu}$, are defined as, e.g., $\bar{n}_{0}^{\nu} = \sum_{\mu \neq \nu} n_{0}^{\mu\mu}$. The bare dispersion enters through $\xi^{\mu\nu}({\bf k}) = \varepsilon^{\mu\nu}({\bf k}) -   \delta^{\mu\nu} \mu_{0}$, where $\epsilon^{\mu\nu}({\bf k}) $ is obtained from the Bloch Hamiltonian entering $ H_{0} + H_{\delta} $.

%-------------------------------------------------------------------------------------
\subsection{RPA correlation function in the stripe-SDW phase}
\label{sec:RPAstripe}
%-------------------------------------------------------------------------------------

\begin{figure*}[t!]
\centering
% 1st =========================================
%-----------------------------------------------------------
\begin{minipage}{0.48\textwidth}
\centering
\flushleft{(a)}
\includegraphics[width=1\columnwidth]{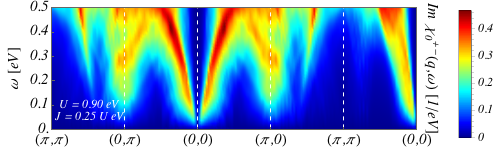}
\end{minipage}
%-----------------------------------------------------------
\begin{minipage}{0.48\textwidth}
\centering
\flushleft{(b)}
\includegraphics[width=1\columnwidth]{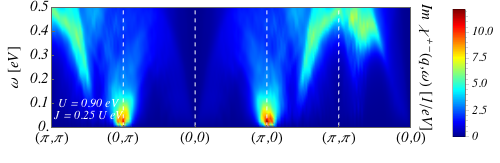}
\end{minipage}
%-----------------------------------------------------------
% 2nd =========================================
%-----------------------------------------------------------
\begin{minipage}{0.48\textwidth}
\centering
\flushleft{(c)}
\includegraphics[width=1\columnwidth]{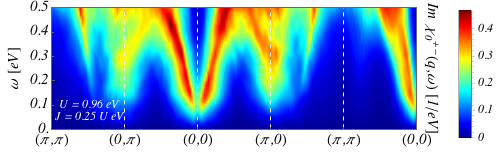}
\end{minipage}
%-----------------------------------------------------------
\begin{minipage}{0.48\textwidth}
\centering
\flushleft{(d)}
\includegraphics[width=1\columnwidth]{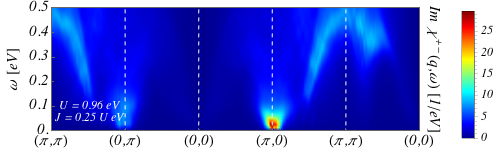}
\end{minipage}
%-----------------------------------------------------------
% 3rd =========================================
%-----------------------------------------------------------
\begin{minipage}{0.48\textwidth}
\centering
\flushleft{(e)}
\includegraphics[width=1\columnwidth]{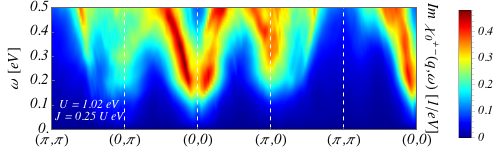}
\end{minipage}
%-----------------------------------------------------------
\begin{minipage}{0.48\textwidth}
\centering
\flushleft{(f)}
\includegraphics[width=1\columnwidth]{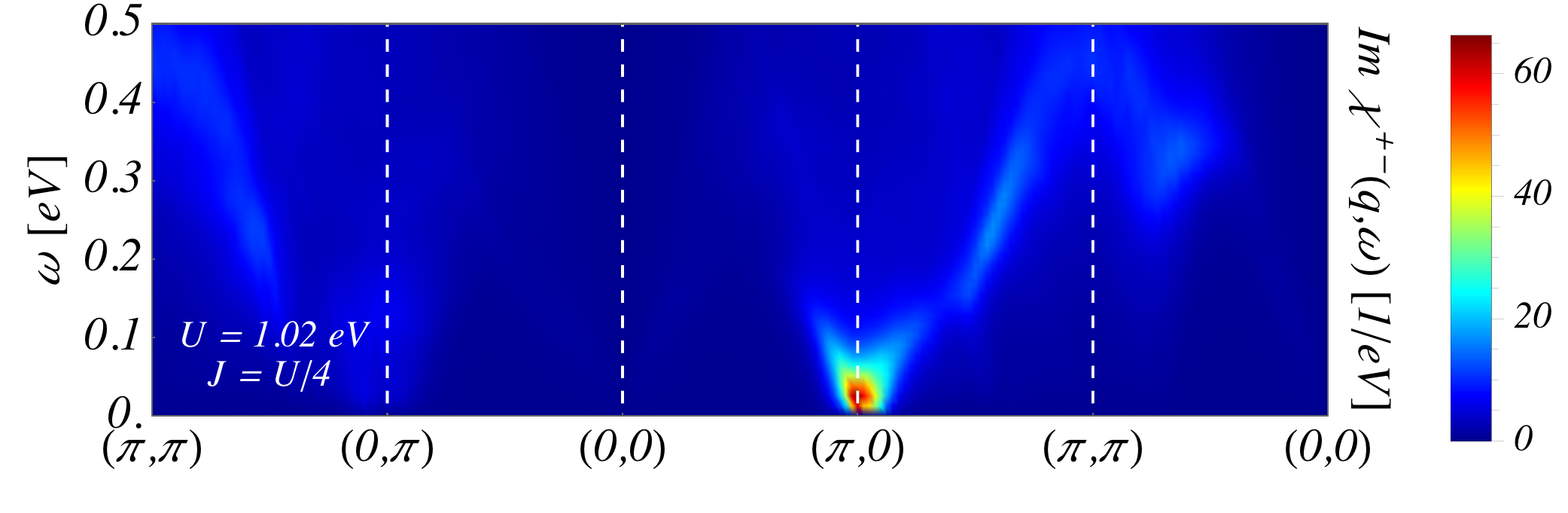}
\end{minipage}
%-----------------------------------------------------------
% 4th =========================================
%-----------------------------------------------------------
\begin{minipage}{0.48\textwidth}
\centering
\flushleft{(g)}
\includegraphics[width=1\columnwidth]{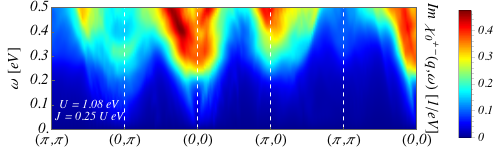}
\end{minipage}
%-----------------------------------------------------------
\begin{minipage}{0.48\textwidth}
\centering
\flushleft{(h)}
\includegraphics[width=1\columnwidth]{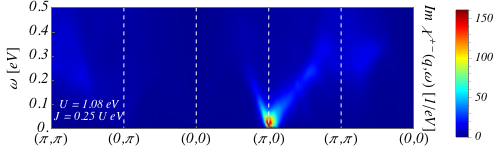}
\end{minipage}
%-----------------------------------------------------------
% 5th =========================================
%-----------------------------------------------------------
\begin{minipage}{0.48\textwidth}
\centering
\flushleft{(i)}
\includegraphics[width=1\columnwidth]{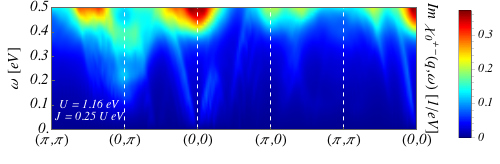}
\end{minipage}
%-----------------------------------------------------------
\begin{minipage}{0.48\textwidth}
\centering
\flushleft{(j)}
\includegraphics[width=1\columnwidth]{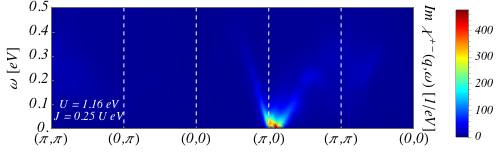}
\end{minipage}
%-----------------------------------------------------------
\caption{High-symmetry cuts of particle-hole spectra (left column) and spin-wave spectra (right column) as a function of $ U $ at $ \delta = 0 $\,eV,  $ T = 0.01 $\,eV, $ J = U/4 $  ($ z = 1 $). The SDW order is just about to form at $ U = 0.90 $\,eV. With increasing $U$, the Goldstone mode gains additional spectral weight and the high-energy excitation mode at $ (\pi,\pi) $ moves down in energy, until the bandwidth of the branch along $ (0,0) - (\pi,0) $ becomes larger than the bandwidth along $ (\pi,0) - (\pi,\pi) $.}
\label{fig:ph_vs_spin}
\end{figure*}

To connect to neutron-scattering experiments, we define the time-ordered (with $\mathcal{T}_{\tau}$ the imaginary-time
ordering operator) Matsubara spin-spin correlation function
\be
\chi^{ij}({\bf q},\mathrm{i}\omega_n) = g^2\int_{0}^{\beta} \! d\tau \,
\mathrm{e}^{\mathrm{i}\omega_n \tau}
\langle \mathcal{T}_{\tau} S^{i}_{{\bf q}}(\tau) S^{j}_{-{\bf q}}(0)\rangle,
\ee
with the $i$th component ($ i = x,y,z $) of the Fourier transformed electron-spin operator for the 1-Fe unit cell given as
\be
S^{i}_{{\bf q}}(\tau) = \frac{1}{\sqrt{\mathcal{N}}}\sum_{{\bf k},\mu,\sigma,\sigma^{\prime}} c_{{\bf k} - {\bf q}\mu\sigma}^{\dagger}(\tau) \frac{\sigma_{\sigma\sigma^{\prime}}^{i}}{2}c_{{\bf k}\mu\sigma^{\prime}}(\tau).
\ee
We note that we work with a coordinate system, where $ z = a $, such that the transverse susceptibility defined below probes spin excitations perpendicular (i.e., in the $xy$ plane of our coordinate system) to the ordered moment along, which is aligned along $ a $. The longitudinal component probes spin excitations parallel to $ z $. Performing an analytic continuation $\mathrm{i}\omega_{n} \to E + \mathrm{i}\eta$, with $\eta \to 0^{+}$, we arrive at the energy and momentum dependent susceptibility tensor $\chi^{ij}({\bf q},E)$. We also note that the Lande factor $ g = 2 $. We work with units where the Bohr magneton $\mu_{\mathrm{B}} = 1$, then the imaginary part $ \chi^{\prime\prime}({\bf q},E) $ has units [1/eV]. Unpolarized neutron-scattering experiments allow to probe the energy and momentum dependent imaginary part of the spin susceptibility. As the spin response is dominated by the transverse (with respect to the $ z = a $ spin-quantization axis) response, we will focus on the transverse channel in the following. The averaged susceptibility is defined as 
\be 
\chi^{\prime\prime}({\bf q},E) & \equiv & \frac{1}{3}\sum_{\alpha,\beta} \mathrm{Im}\chi^{\alpha\beta}({\bf q},E) \\
& \simeq &  \frac{1}{3}\mathrm{Im}\left[ \chi^{xx}({\bf q},E) + \chi^{yy}({\bf q},E) \right] \\
& = &  \frac{4}{3}\mathrm{Im}\left[ \chi^{+-}({\bf q},E) + \chi^{-+}({\bf q},E) \right] \\
& = &  \frac{8}{3}\mathrm{Im}\left[ \chi^{+-}({\bf q},E) \right],
\ee
where we neglected the longitudinal $\chi^{zz}({\bf q},E)$ contribution that is gapped in the collinear AF ordered state with moments aligned along $z$. The Matsubara correlation function corresponding to the transverse susceptibility is given by
\be
\chi^{+-}({\bf q},\mathrm{i}\omega_n) = \frac{1}{2}\int_{0}^{\beta} \! d\tau \,
\mathrm{e}^{\mathrm{i}\omega_n \tau}
\langle \mathcal{T}_{\tau} S^{+}_{{\bf q}}(\tau) S^{-}_{-{\bf q}}(0)\rangle,
\ee
with the electron spin-raising and lowering operators
\be
S^{+}_{{\bf q}}(\tau) = \frac{1}{\sqrt{\mathcal{N}}}\sum_{\mu,{\bf k}} c_{{\bf k} - {\bf q}\mu\uparrow}^{\dagger}(\tau) c_{{\bf k}\mu\downarrow}(\tau),\quad
S^{-}_{{\bf q}}(\tau) = \frac{1}{\sqrt{\mathcal{N}}}\sum_{\mu,{\bf k}} c_{{\bf k} - {\bf q}\mu\downarrow}^{\dagger}(\tau) c_{{\bf k}\mu\uparrow}(\tau).
\ee
In the following we briefly describe the RPA formalism to extract the spin excitations of the SDW phase. We note that for a vanishing mean-field magnetic order-parameter, the formalism turns into a glorified version of the standard multiorbital RPA in the paramagentic state. It proves useful to introduce a correlation function with two independent momenta ${\bf q}$, ${\bf q^{\prime}}$ as
\be
[\chi^{+-}]^{\mu_1 \mu_2}_{\mu_3 \mu_4}({\bf q},{\bf q^{\prime}},\mathrm{i}\omega_n)=
\frac{1}{\mathcal{N}}\int_{0}^{\beta} \! d\tau \, \mathrm{e}^{\mathrm{i}\omega_n \tau}\sum_{{\bf k},{\bf k}^{\prime}}
\langle \mathcal{T}_{\tau} 
c_{{\bf k} - {\bf q}\mu_1\uparrow}^{\dagger}(\tau) 
c_{{\bf k}\mu_2\downarrow}(\tau)
c_{{\bf k} + {\bf q^{\prime}}\mu_3\downarrow}^{\dagger}(0) 
c_{{\bf k}\mu_4\uparrow}(0)
\rangle.
\ee
To take into account virtual Umklapp scattering processes in the computation of the RPA susceptibility, we then define a generalized correlation function
\be 
[\chi_{l,l^{\prime}}^{+-}]^{\mu_1 \mu_2}_{\mu_3 \mu_4}({\bf q},\mathrm{i}\omega_n) \equiv
[\chi^{+-}]^{\mu_1 \mu_2}_{\mu_3 \mu_4}
({\bf q} + {\bf Q}_{l},{\bf q} + {\bf Q}_{l^{\prime}},\mathrm{i}\omega_n)
\ee
where for $l=0$, ${\bf Q}_{0} = (0,0)$, and for $l = 1$, ${\bf Q}_{1} = (\pi,0)$. The correlation function thus becomes a $2 \times 2$ matrix in `Umklapp space'. In the presence of stripe-SDW, the orbital-to-band matrix elements entering the components of the bare matrix-valued particle-hole propagator become
\be 
[\mathcal{M}_{n_1,n_2;l,l^{\prime}}({\bf k},{\bf q})]^{\mu_1 \mu_2}_{\mu_3 \mu_4}=
\sum_{[l_1,l_2,l_3,l_4]_{l,l^{\prime}}}
\mathcal{U}_{{\mu_1l_1,n_1}}^{\ast}({\bf k} - {\bf q},\uparrow)
\mathcal{U}_{{\mu_2l_2,n_2}}({\bf k},\downarrow)
\mathcal{U}_{{\mu_3l_3,n_2}}^{\ast}({\bf k},\downarrow)
\mathcal{U}_{{\mu_4l_4,n_1}}({\bf k} - {\bf q},\uparrow),
\ee
where $\sum_{[l_1,l_2,l_3,l_4]_{l,l^{\prime}}}$ denotes a restricted sum
over $l$-index tuples contributing to the $l$,$l^{\prime}$ component of the
particle-hole propagator. The non-interacting correlation function can then be written as
\be 
[\chi_{0;l,l^{\prime}}^{+-}]^{\mu_1 \mu_2}_{\mu_3 \mu_4}({\bf q},\mathrm{i}\omega_n) =
-\frac{1}{\mathcal{N}}\sum_{{\bf k},n_1,n_2}^{\prime}
[\mathcal{M}_{n_1,n_2;l,l^{\prime}}({\bf k},{\bf q})]^{\mu_1 \mu_2}_{\mu_3 \mu_4}
\frac{f(E_{n_1}({\bf k}-{\bf q})) - f(E_{n_2}({\bf k}))}{\mathrm{i}\omega_n + E_{n_1}({\bf k}-{\bf q}) - E_{n_2}({\bf k})},
\ee
and the prime on the sum denotes a ${\bf k}$ summation over the corresponding rBZ. The RPA equation for the transverse correlation function in the magnetic phase reads
\be 
[\chi^{+-}_{l,l^{\prime}}]^{\mu_1 \mu_2}_{\mu_3 \mu_4}({\bf q},\mathrm{i},\omega_n) =
[\chi_{0;l,l^{\prime}}^{+-}]^{\mu_1 \mu_2}_{\mu_3 \mu_4}({\bf q},\mathrm{i}\omega_n) +
[\chi_{0;l,\bar{l}}^{+-}]^{\mu_1 \mu_2}_{\nu_1 \nu_2}({\bf q},\mathrm{i}\omega_n)
[U]^{\nu_1 \nu_2}_{\nu_3 \nu_4}
[\chi^{+-}_{\bar{l},l^{\prime}}]^{\nu_3 \nu_4}_{\mu_3 \mu_4}({\bf q},\mathrm{i}\omega_n),
\ee
where the vertex $[U]^{\nu_1 \nu_2}_{\nu_3 \nu_4}$ that describes the orbitally dependent interaction of electrons with spin fluctuations is given by
\be
[U]^{\mu\mu}_{\mu\mu} = U,
\quad
[U]^{\nu\mu}_{\mu\nu} = U^{\prime},
\quad
[U]^{\nu\nu}_{\mu\mu} = J,
\quad
[U]^{\mu\nu}_{\mu\nu} = J^{\prime},
\quad\text{with}\,\mu \neq \nu.
\ee
In a condensed matrix-notation following the conventions given above, we can write the solution to this linear equation as
\be\label{eq:RPA}
[\chi^{+-}]({\bf q},\mathrm{i}\omega_{n})  = \left[ \mathbbm{1} - [\chi_{0}^{+-}]({\bf q},\mathrm{i}\omega_{n}) [U] \right]^{-1} [\chi_{0}^{+-}]({\bf q},\mathrm{i}\omega_{n}).
\ee
After analytic continuation, the physical transverse susceptibility in the RPA approximation is now given by
\be 
\chi^{+-}({\bf q},E) = \frac{1}{2}\sum_{\mu,\nu}[\chi_{0,0}^{+-}]_{\nu\nu}^{\mu\mu}({\bf q},E).
\ee
To account for the effect of moderate correlation effects, we include a phenomenological dynamical self-energy
of the form $\Sigma^{\mu\nu}({\bf k},\mathrm{i}\omega_{n}) = ( 1 -  1/z ) \mathrm{i}\omega_{n}\delta_{\mu\nu} $, $ 0 < z \leq 1 $ in the electronic Greens function. The $z$-factor simulates both a reduced quasiparticle weight as well as a uniform band-renormalization of the quasiparticle states. We argue here, as is done in the microscopic justification of Fermi liquid theory, that the vertex that describes the coupling of quasiparticles to an external magnetic field is renormalized by a corresponding factor $z^{-1}$, originating from screening due to the incoherent part of the electronic spectrum~\cite{leggett1965}. This gives rise to another multiplicative factor $ z^{-2} $ for the susceptibility. One can show that $ \chi^{\prime\prime}({\bf q},E) $ can be represented as
\be 
\chi^{\prime\prime}({\bf q},E) =  \frac{8}{3 z} \mathrm{Im}\left[\tilde{\chi}^{+-}({\bf q},E/z)\right],
\ee
where $ \tilde{\chi}^{+-}({\bf q},\tilde{E}) $ is a susceptibility obtained for  $ \tilde{T} = T/z$, $ \tilde{E} = E/z$ and $ [\tilde{U}] = z [U] $. This can be read as follows: for a fixed set of parameters $ \tilde{T} $, $ [\tilde{U}] $ we compute the RPA approximation $ [\tilde{\chi}^{+-}]({\bf q},\tilde{E}) $ as a function of $ \tilde{E} $. We can then use $ z $ as a fitting parameter to match the experimentally observed bandwidth for magnetic excitations. At the same time, however, the spectral weight is multiplied by a factor $ 1/z $. Once the fitting-parameter $ z $ has been fixed by comparison to experiment, we can relate the temperature and interaction parameters from our calculation to the `true' parameters $ T $ and $ [U] $. For the calculations in the
magnetic state presented in the main text, we used $\tilde{T} = 0.01$\,eV and $ \tilde{U} = 1.02 $\, eV (with $ \tilde{J} = \tilde{U}/4 $). The physical values for a given $z$-factor are then obtained as $ T = z \tilde{T}$, $ U = \tilde{U}/z $. While we thereby can achieve agreement with the shape of the experimentally obtained spin-wave spectra, the spectral weight of our renormalized RPA calculation still comes out too small. The evolution of the spin-excitation spectrum with increasing interaction for vanishing orbital splitting is shown (with $ z = 1$) in Fig.~\ref{fig:ph_vs_spin}. In Fig.~\ref{fig:ph_vs_spin}, we also show the evolution of the spectral weight of the particle-hole continuum, that is renormalized due to the mean-field self-energy describing the SDW order. As the particle-hole continuum gives rise to Landau damping of collective spin excitations, the particle-hole spectra explain the stronger damping of the $ (0,0) - (\pi,0) $ branch as compared to the $ (\pi,0) - (\pi,\pi) $ branch for weak interactions. The gapping of particle-hole excitations with increasing interactions eventually leads to more well-developed spinwave branches also along the $ (0,0) - (\pi,0) $ direction.

\begin{figure*}[t!]
\centering
%-----------------------------------------------------------
\begin{minipage}{0.3\textwidth}
\centering
\flushleft{(a)}
\includegraphics[width=1\columnwidth]{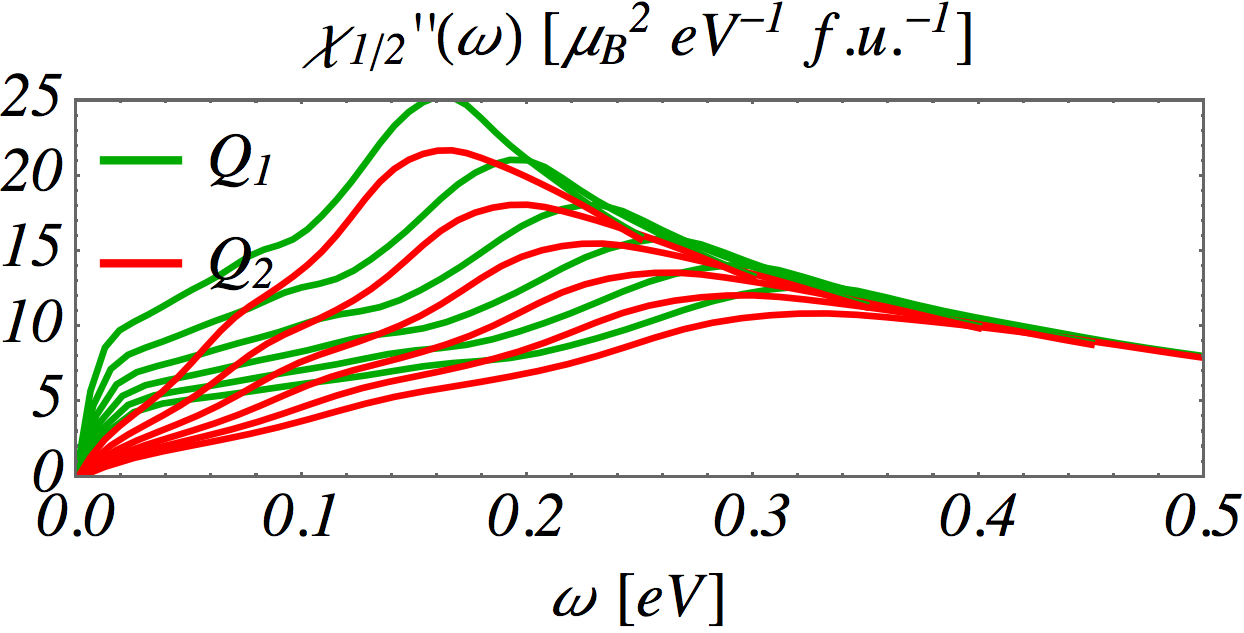}
\end{minipage}
%-----------------------------------------------------------
\begin{minipage}{0.3\textwidth}
\centering
\flushleft{(b)}
\includegraphics[width=1\columnwidth]{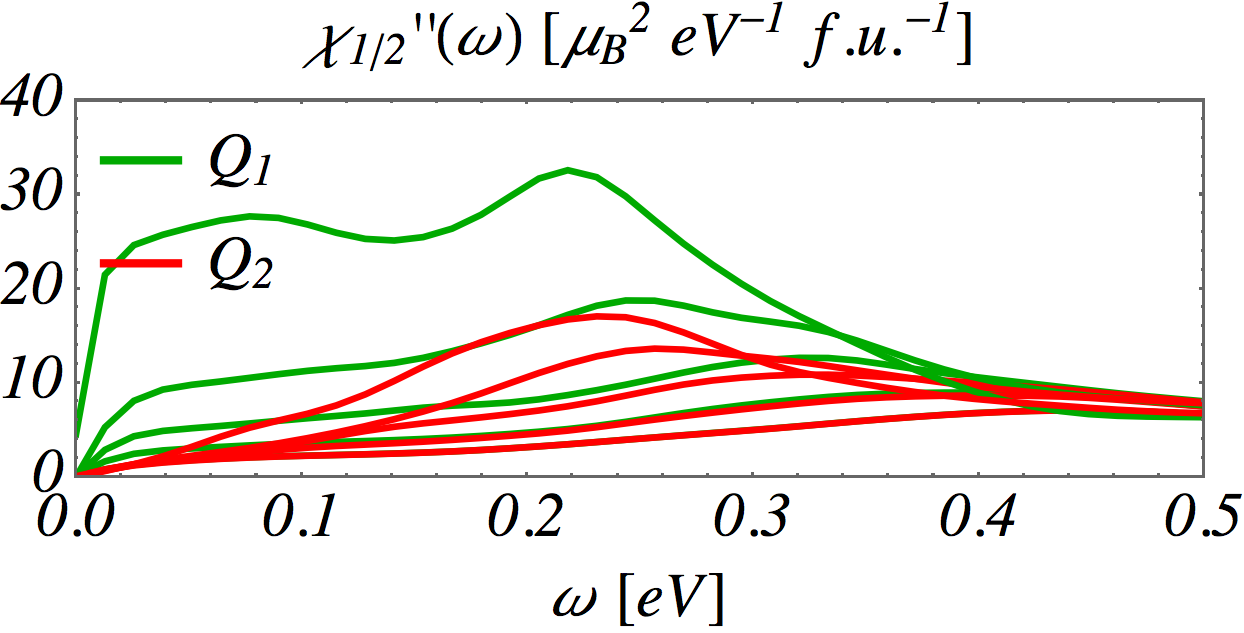}
\end{minipage}
%-----------------------------------------------------------
\begin{minipage}{0.3\textwidth}
\centering
\flushleft{(c)}
\includegraphics[width=1\columnwidth]{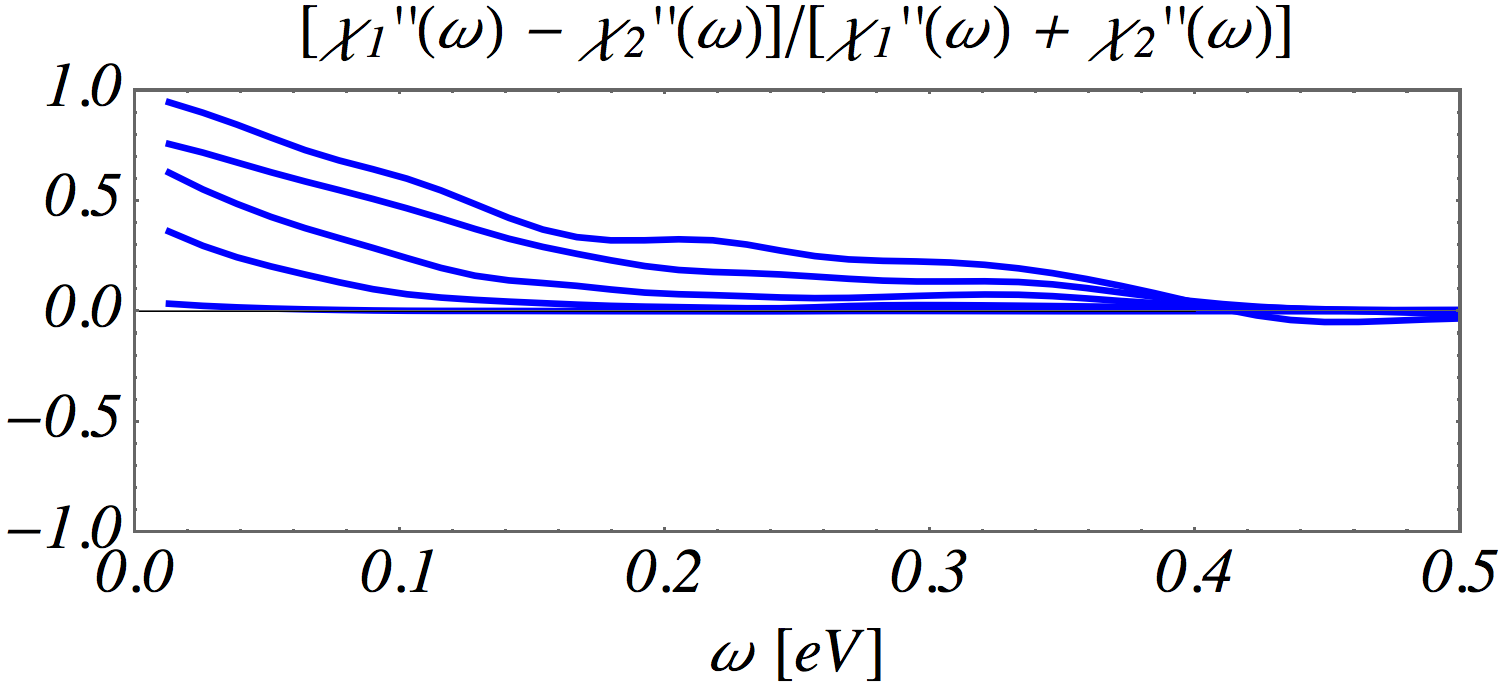}
\end{minipage}
%-----------------------------------------------------------
\caption{(a) Dependence of the local susceptibilities as function of energy on the renormalization parameter $z$ for effective interaction parameters $ U = 1.02/z \, $eV, $ J = U/4$. The parameter $ z $ is decreased in steps of $\Delta z = 0.1$ from $z = 1$ down to $ z = 0.5 $. The lowermost curves correspond to $ z = 1 $. We note, that as we have not produced data for $\omega > 0.5 $eV, the curves are actually cut off at $0.5 z \, $eV. As $ z $ decreases, the peak in the local susceptibilities moves to lower energies and the intensity increases. (b) Dependence of the local susceptibilities as function of energy on the interaction parameter $ U $ with $ J = U/4 $ and $ z = 1$. The five sets of curves correspond to the spectra shown in Fig.~\ref{fig:ph_vs_spin}. The intensity increases monotoneously with interaction strength. (c) Evolution of the anisotropy for the spectra shown in Fig.~\ref{fig:ph_vs_spin} with interaction paramter $U$. The lowest curve with almost vanishing anisotropy correspodns to $ U = 0.90 \,$eV. At low energies, the anisotropy increases monotoneously with increasing interaction strength.}
\label{fig:chi_par_dep}
\end{figure*}

To explore the parameter dependence of our results, we have looked at how the local susceptibilities change, as the renormalization parameter $ z $ changes. We plotted the results in Fig.~\ref{fig:chi_par_dep}(a). The parameter $ z $ is decreased in steps of $\Delta z = 0.1$ from $z = 1$ down to $ z = 0.5 $. The lowermost pair of red and green curves in Fig.~\ref{fig:chi_par_dep}(a) corresponds to $ z = 1 $. As $ z $ decreases, the peak in the local susceptibilities moves to lower energies and the intensity increases. As the anisotropy is not affected by prefactors, we did not replot it for different choices of $ z $. While also values $ z < 0.7 $ might seem compatible with the experimental data, the agreement of the shape of the spectral distribution (in other words, the spin-wave velocities) and the experimental data would worsen. In Fig.~\ref{fig:chi_par_dep}(b) we collect the local susceptibilities as they result for the spectra shown in Fig.~\ref{fig:ph_vs_spin}. In this way, we can explore the effect of increasing interaction strength $ U $ (with fixed ratio $ J/U = 1/4$) on the local susceptibilities. The corresponding anisotropy is displayed in Fig.~\ref{fig:chi_par_dep}(c). While interactions $ U > 1.02 \, $ eV yield both larger susceptibilities and anisotropy, the agreement between the shape of the spectra (see Fig.~\ref{fig:ph_vs_spin}) worsens. In particular, for $ U = 1.16 \, $eV the spinwave spectrum has undergone a crossover to a situation where the bandwidth of the dispersive $ (0,0) - (\pi,0) $ branch becomes larger than that of the $ (\pi,0) - (\pi,\pi) $ branch. The latter situation is in conflict with the experimental data. Therefore, the search for interaction parameters that give rise to an acceptable description of the spinwave dispersion is limited to values $ U < 1.16 \, $eV.

\begin{figure*}[t!]
\centering
%-----------------------------------------------------------
\begin{minipage}{0.48\textwidth}
\centering
\flushleft{(a)}
\includegraphics[width=1\columnwidth]{./spinwaves_T0p01_n6p00_U1p02_J0p25U}
\end{minipage}
%-----------------------------------------------------------
\begin{minipage}{0.48\textwidth}
\centering
\flushleft{(b)}
\includegraphics[width=1\columnwidth]{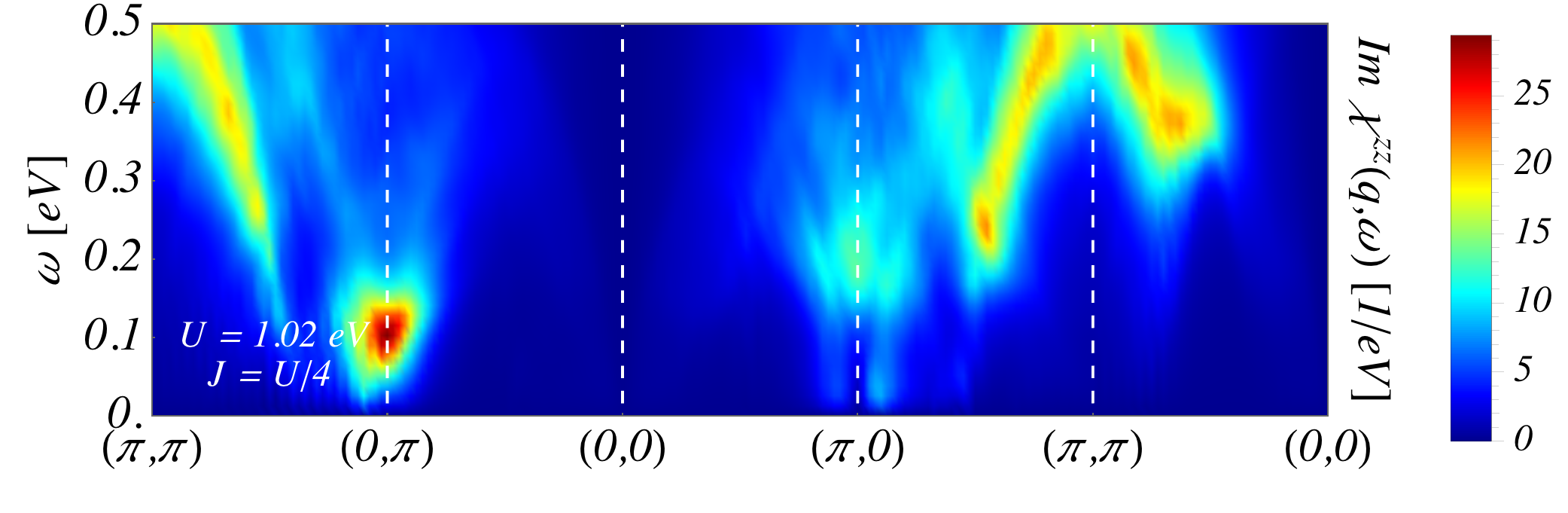}
\end{minipage}
%-----------------------------------------------------------
\caption{Comparison of high-symmetry cuts through the spectral weight distributions of collective magnetic excitations in (a) transverse and (b) longitudinal channels for $ U = 1.02 \, $eV and $ J = U/4 $ (with $ z = 1$). We note that due to our conventions, the transverse component enters with a relative weight-factor 8 into the total spectral weight, compared to the longitudinal contribution.}
\label{fig:trans_vs_long}
\end{figure*}

\begin{figure*}[t!]
\centering
%-----------------------------------------------------------
\begin{minipage}{0.40\textwidth}
\centering
\flushleft{(a)}
\includegraphics[width=0.96\columnwidth]{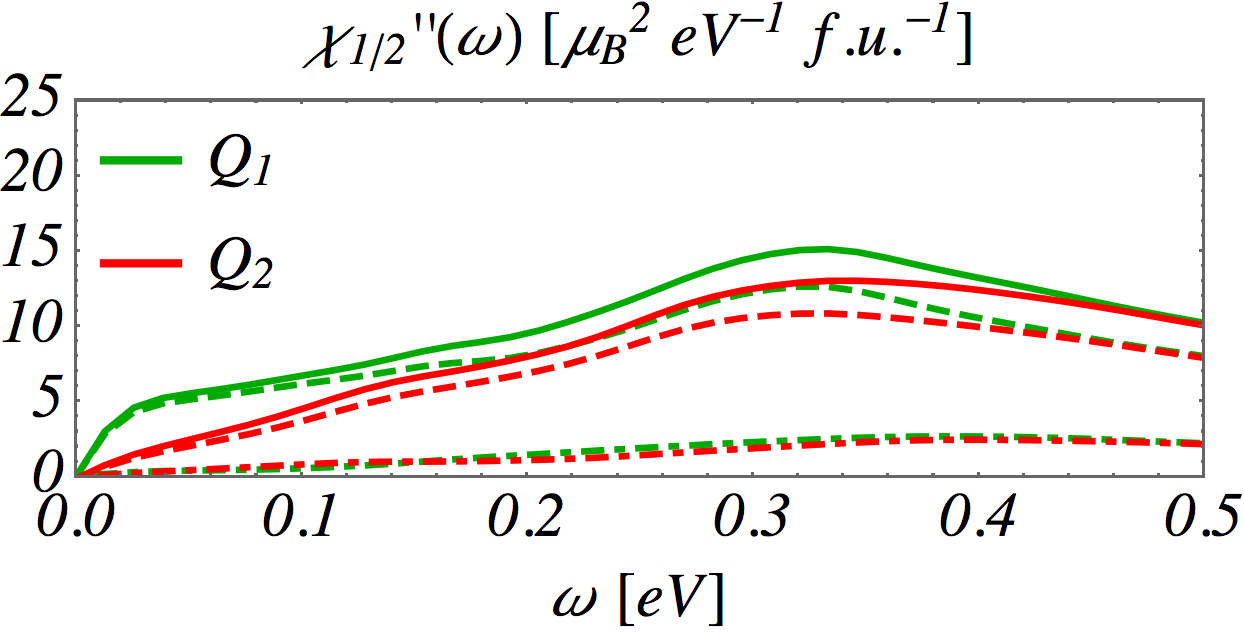}
\end{minipage}
%-----------------------------------------------------------
\hspace{1cm}
%-----------------------------------------------------------
\begin{minipage}{0.40\textwidth}
\centering
\flushleft{(b)}
\includegraphics[width=1\columnwidth]{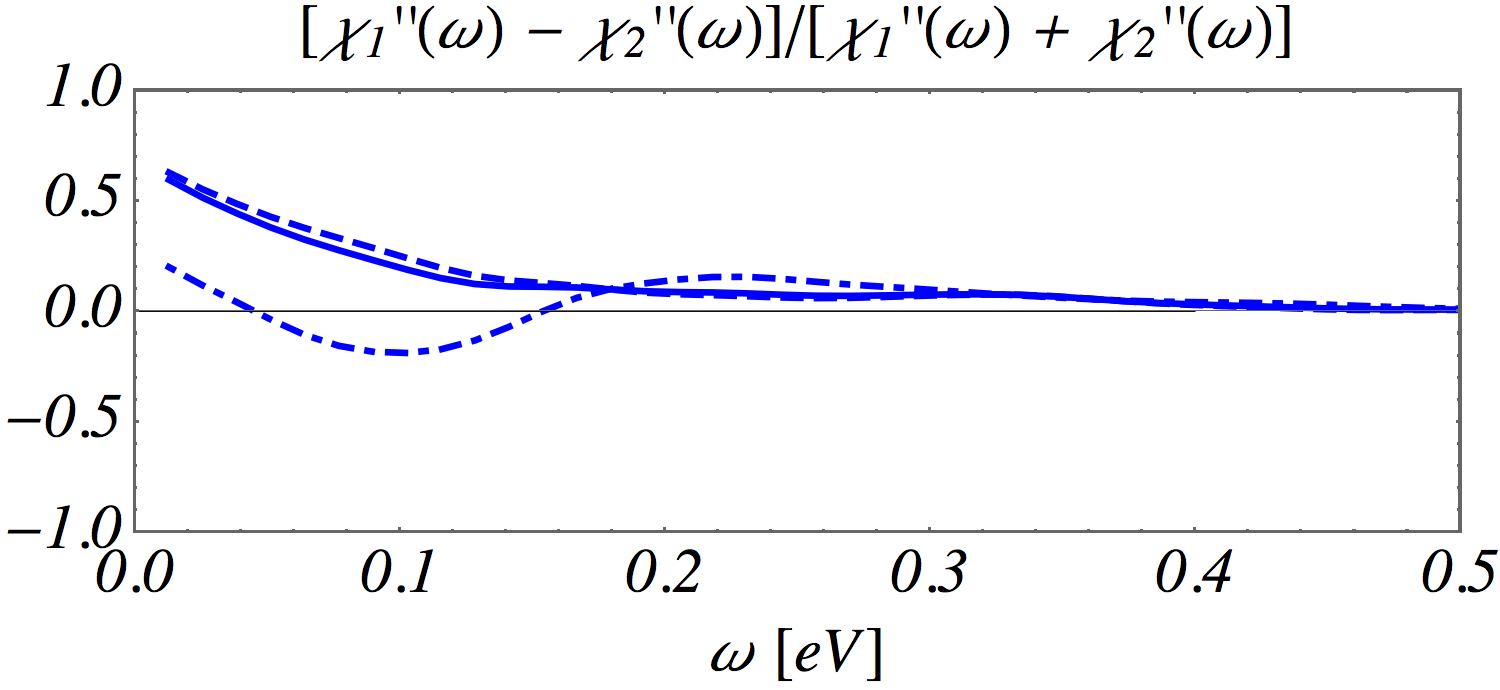}
\end{minipage}
%-----------------------------------------------------------
\caption{(a) Local susceptibilities around momenta ${\bf Q}_{1} = (\pi,0)$ and ${\bf Q}_{2} = (0,\pi)$ as a function of energy, calculated with transverse (dashed), longitudinal (dot-dashed) and the weighted sum of transverse and longitudinal (solid) contributions. The interaction parameters are $ U = 1.02 \, $eV and $ J = U/4 $ with renormalization parameter $ z = 1 $. (b) The anisotropy curves corresponding to the local susceptibilities shown in (a).}
\label{fig:chi_local_long}
\end{figure*}

Taking into account the longitudinal component, the formula for $ \chi^{\prime\prime}({\bf q},E) $ is modified as
\be 
\chi^{\prime\prime}({\bf q},E) =  \frac{1}{3 z} 
\mathrm{Im}
\left[
8 \tilde{\chi}^{+-}({\bf q},E/z) + \tilde{\chi}^{zz}({\bf q},E/z)
\right],
\ee
where the different numerical prefactors for transverse and longitudinal contributions are a consequence of our conventions. We note that in a paramagnetic state (neglecting spin-orbit coupling) we would have $ \tilde{\chi}^{+-}({\bf q},E/z) = \frac{1}{4} \tilde{\chi}^{zz}({\bf q},E/z) $, such that $\chi^{\prime\prime}({\bf q},E) =  \frac{1}{z} \tilde{\chi}^{zz}({\bf q},E/z) $. The RPA-result for $ \tilde{\chi}^{zz}({\bf q},E/z) $ is obtained from an equation analogous to Eq.~\eqref{eq:RPA}, where the both the bare susceptibility and the interaction vertex have to be adapted to the case of longitudinal excitations. In Fig.~\ref{fig:trans_vs_long}, we compare the transverse and longitudinal excitation spectra (with $ z = 1$). The longitudinal excitation spectrum shows a pronounced spin gap at the ordering vector $(\pi,0)$. A dispersive high-energy branch emerges only at higher energies. Most of the spectral weight in the longitudinal channel is concentrated in a non-dispersive, gapped excitation at $(0,\pi)$. The weight in the transverse channel at momentum $(0,\pi)$ in the corresponding energy range is actually even larger. We note that due to our conventions, the transverse component enters with a relative weight-factor 8 into the total spectral weight, compared to the longitudinal contribution. It is clear, that the low-energy anisotropy of the longitudinal channel is opposite to the anisotropy in the transverse channel (where we refer to the anisotropy $\psi({\bf q},E)$ of the local susceptibilities $ \chi_{1}^{\prime\prime}(E) $ and $ \chi_{2}^{\prime\prime}(E) $ as defined in the main text). Considering the total anisotropy, the inclusion of the longitudinal contribution in fact slightly diminishes the anisotropy at low energies, see Fig.~\ref{fig:chi_local_long}(b). Quantitatively, however, the effect is marginal. We therefore neglect the longitudinal component in our modelling of the anisotropy as discussed in the main text. Considering the anisotropy of the longitudinal contribution alone, it turns negative once the energy passes the excitation gap at $(0,\pi)$, which is much smaller than the spin gap at $(\pi,0)$.

\newpage
\clearpage

\section{Sample and neutron scattering experiment}
\subsection{Sample, detwinning device and experimental setup}

\begin{figure}[htbp]
\includegraphics[width=14cm]{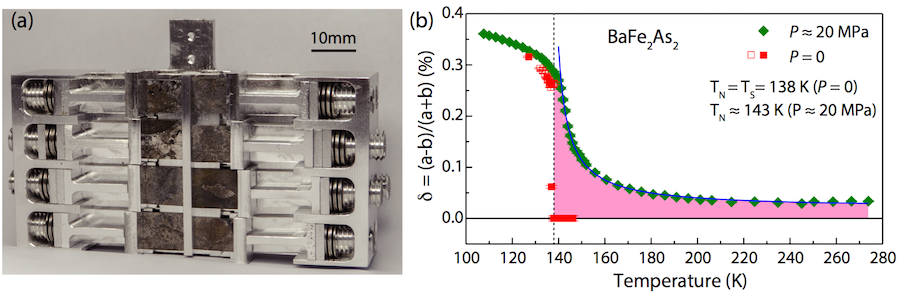}
\caption{ (a) The mechanical device used to simultaneously detwin 16 pieces of large {\BFA} single crystals (with a total mass of $6.4$ grams). The horizontal black line marks a scale of 10 mm. (b) Orthorhombic lattice distortions of {\BFA} measured under zero pressure (red open squares) and $\sim20$MPa (green filled diamonds), measured using Larmor diffraction at TRISP (Three axes spin echo spectrometer), MLZ, Germany . The pink area marks the pressure-induced orthorhombic lattice distortions above $T_S$.}
\label{fig.F1}
\end{figure} 

The {\BFA} single crystals used in the present study were grown using self-flux method as described elsewhere \cite{yanchao}. To prepare square-shaped {\BFA} crystals, we selected more than 20 pieces of large, flat {\BFA} single crystals and determined their tetragonal [1,1,0] orientation using a Laue camera. The selected crystals were cut along the tetragonal [1,1,0] and [1,-1,0] directions by a high-precision wire saw. The well-cut crystals were inserted into the slots of the detwinning device as shown in Fig. \ref{fig.F1} (a). Uniaxial pressures were applied by pressing the spring washers at two ends \cite{science14}. Depending on the sectional areas of the crystals, the applied uniaxial pressures ranges from $\sim 12$ MPa to $\sim 22$ MPa. Fig. \ref{fig.F1}(b) shows the orthorhombic lattice distortion of {\BFA} under zero pressure and $20$MPa \cite{Lu16PRB}. The uniaxial pressure induces an orthorhombic lattice distortion in temperature range above $T_S$, but the pressure-induced distortion decreases drastically below $T_S$ \cite{Lu16PRB}.

For our time-of-flight neutron scattering experiments at MERLIN spectrometer, incident beam was set to be perpendicular to the sample surface and most area of the device and the spring washers were covered by neutron absorbing B$_4$C. The horizontal slit 
of neutron beam was set to 27 mm, to reduce the background arising from incoherent scattering and multiple scattering. At $T=7$ K, the incident neutron energies were set to $E_i=80, 250, 450$ meV to cover spin waves in the energy range of $E\approx 0-320$ meV. $E_i\approx80, 160, 250$ meV were used for the measurements across $T_N$. We define the wave vector \textbf{Q} in three-dimensional reciprocal space in \AA$^{-1}$ as ${\bf Q}=H {\bf a^\ast}+K{\bf b^\ast}+L{\bf c^\ast}$, where $H$, $K$, and $L$ are Miller indices and ${\bf a^\ast}=\hat{{\bf a}}2\pi/a, {\bf b^\ast}=\hat{{\bf b}} 2\pi/b, {\bf c^\ast}=\hat{{\bf c}}2\pi/c$ are reciprocal lattice units (r.l.u.) [Fig. 1(b)]. In the low-temperature AF orthorhombic phase of {\BFA}, $a\approx 5.62$ \AA, $b\approx 5.57$ \AA, and $c\approx 12.94$ {\AA}. 

\subsection{Local dynamic susceptibility $\chi''(E)$ and magnon dispersion}

Dynamic susceptibility $\chi''(\mathbf{Q}, E)$ can be calculated directly from background-substracted magnetic scattering signal obtained on a time-of-flight spectrometer. By comparing it to the incoherent scattering of a standard vanadium sample, $\chi''(\mathbf{Q}, E)$ can be normalized to absolute intensity (with the unit of barn$^{-1}$ sr$^{-1}$eV$^{-1}$). Note that neutron scattering only probes transverse response that perpendicular to the momentum transfer $\mathbf{Q}$, making the measurements of all components for $\chi''(\mathbf{Q}, E)$ complicated. In previous studies concerning the dynamic susceptibility of iron pnictides, $\chi''(\mathbf{Q}, E)$ was usually assumed to be isotropic in spin space (though it is inaccurate especially for low energy part) \cite{leland11, haydenprb, mengshunphys}. For an isotropic magnetic system, its magnetic scattering cross section can be written as:
\begin{eqnarray}\label{1formula1}
\frac{d^2\sigma}{d{\mathbf{\Omega}}dE}=\frac{2(\gamma r_e)^2}{\pi g^2 \mu^2_B}\frac{k_f}{k_i}\big|F(\mathbf{Q})\big|^2\frac{\chi''(\mathbf{q}, E)}{1 - {\rm exp}(-\hbar\omega / k_B T)}
\end{eqnarray}
where ($\gamma r_e$)$^2 = 0.2905 ~\rm{barn~sr}^{-1}$, ${\bf k}_i$, ${\bf k}_f$ are incident and outgoing wave vectors, respectively, $\mathbf{q}$ is reduced wave vector and $F\rm{(}\mathbf{Q}\rm{)}$ the magnetic form factor for Fe$^{2+}$. The raw data shown in Fig. 2 (main text) is $\frac{d^2\sigma}{d{\mathbf{\Omega}}dE}\frac{k_i}{k_f}$ ($S(\mathbf{q}, E)$). To calculate $\chi''(\mathbf{Q}, E)$ or compare the data with the calculation, the other constant coefficients ($\frac{2(\gamma r_e)^2}{\pi g^2 \mu^2_B}=\frac{0.04623}{\mu^2_B}$), magnetic form factor and bose factor ($\approx1$ for all the data shown collected at $T=7$K) needs to be considered.

\begin{figure}[htbp]
\includegraphics[width=14cm]{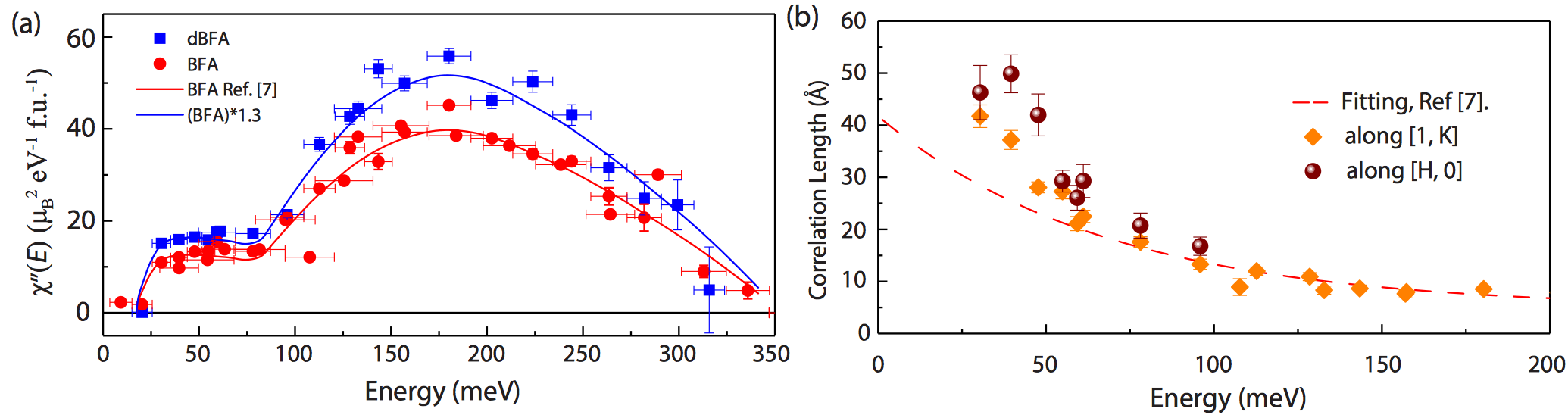}
\caption{ Comparison of local susceptibility and magnetic correlation length between detwinned {\BFA} (dBFA) and twinned {\BFA} (BFA) \cite{leland11}. (a) Local susceptibility $\chi''(E)$ of detwinned (blue squares) and twinned (red dots) {\BFA}. The blue and red solid lines are guides to the eye. The blue line is overall $30\%$ higher than the red line. The horizontal error bars indicate the energy ranges for the intensity integration for calculating $\chi''(E)$. The vertical error bars come  from the uncertainty of the scattered neutrons, taking into account the propagation of the uncertainty for the calculation of $\chi''(E)$. (b) Energy-dependent magnetic correlation length along $[H, 0]$ and $[1, K]$ directions. The red dashed line denotes the magnetic correlation length for twinned {\BFA} reported in Ref. \cite{leland11}. The error bars mark the 1$\sigma$ confidential interval for the fitting of the correlation length using Gaussians.}
\label{fig.F2}
\end{figure}

\begin{figure*}[htbp]
\includegraphics[width=12cm]{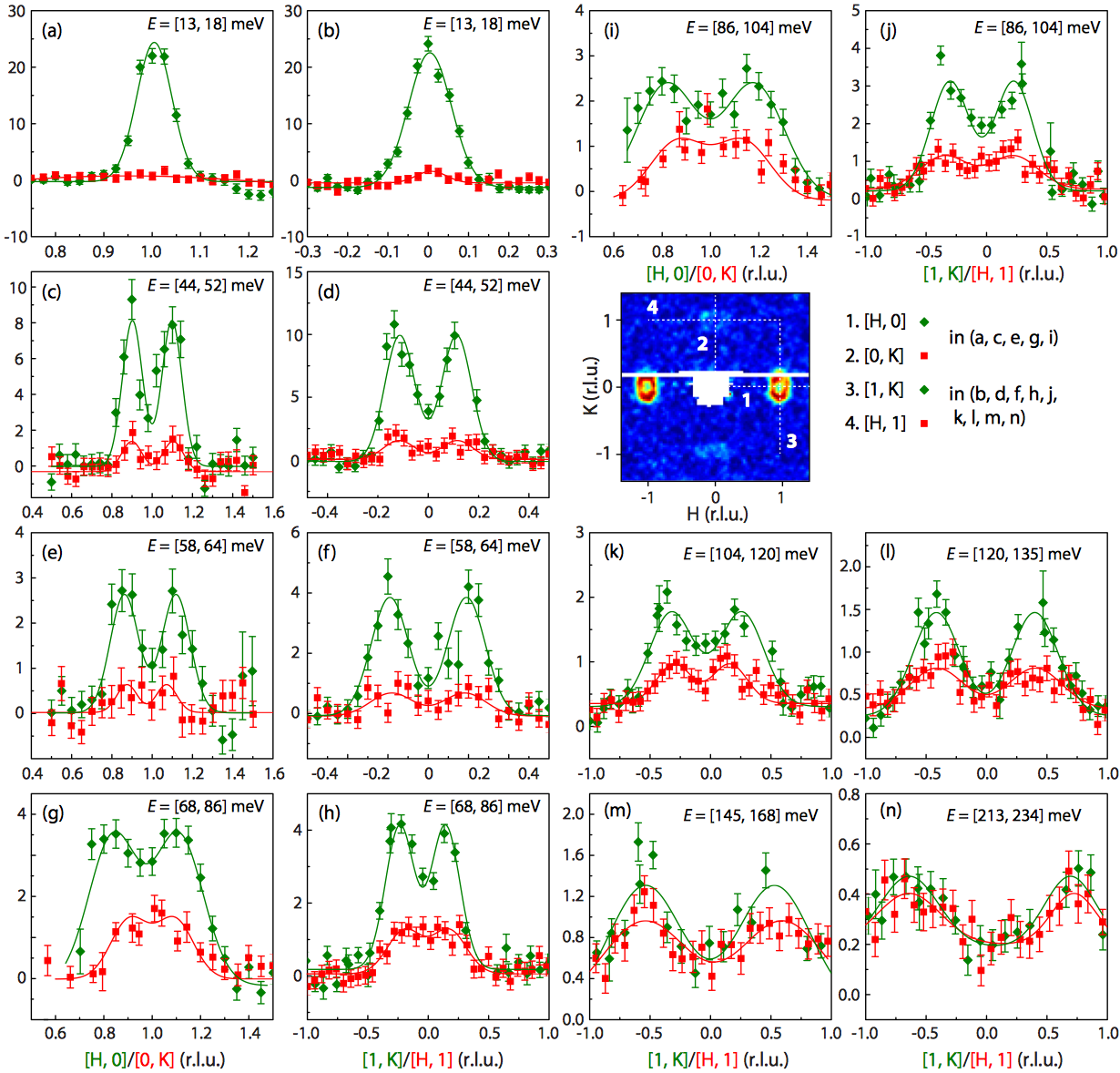}
\caption{ One-dimensional constant-energy cuts along transverse and longitudinal directions across $(1, 0)$ and $(0, 1)$. The directions of the cuts are marked by the dashed lines in the colormap. For the energies from $13$ meV to $104$ meV ((a) to (j)), both transverse ($[1, K]$ and $[H, 1]$) and longitudinal ($[H, 0]$ and $[0, K]$) cuts are present. For energies above 104 meV, only transverse cuts are shown.}
\label{fig.F3}
\end{figure*}

Fig. \ref{fig.F2} summarizes the (1, 0)/(0, 1) averaged absolute local susceptibility and energy-dependent magnetic correlation length of detwinned {\BFA}. 
Fig. \ref{fig.F2}(a) reveals that the detwinned {\BFA} sample shows $30\%$ stronger local susceptibility $\chi''(\omega)$ than that in twinned {\BFA} \cite{leland11}. 
While this is within the errors of our absolute intensity measurements, the difference may be attributed to several sources. First, the counting efficiencies of MAPS (where twinned {\BFA} was measured \cite{leland11}) and MERLIN (where detwinned {\BFA} was measured) spectrometers may be slightly different, depending on the accuracy of detector-efficiency calibration and data vanadium standard normalization. 
Second, the single crystals used for Ref. \cite{leland11} contains some flux (while the well-cut single crystals for detwinning are much cleaner), resulting in slight underestimate of the magnitude of $\chi''(\omega)$. Fig. \ref{fig.F2}(b) shows the energy-dependent magnetic correlation length of detwinned {\BFA}. The correlation length for $E\gtrsim 60$ meV is consistent with that for previous results [red dashed line in Fig. \ref{fig.F2}(b)] \cite{leland11}, while that below $\sim 60$ meV is much larger \cite{leland11}, indicating higher quality of our well-cut crystals. As shown in Fig. 2 (main text) and Fig. \ref{fig.F3}, an elliptical spin wave ring is observed in constant energy slices of the detwinned {\BFA} sample but absent in previous report \cite{leland11}.

\begin{figure*}[htbp]
\includegraphics[width=12cm]{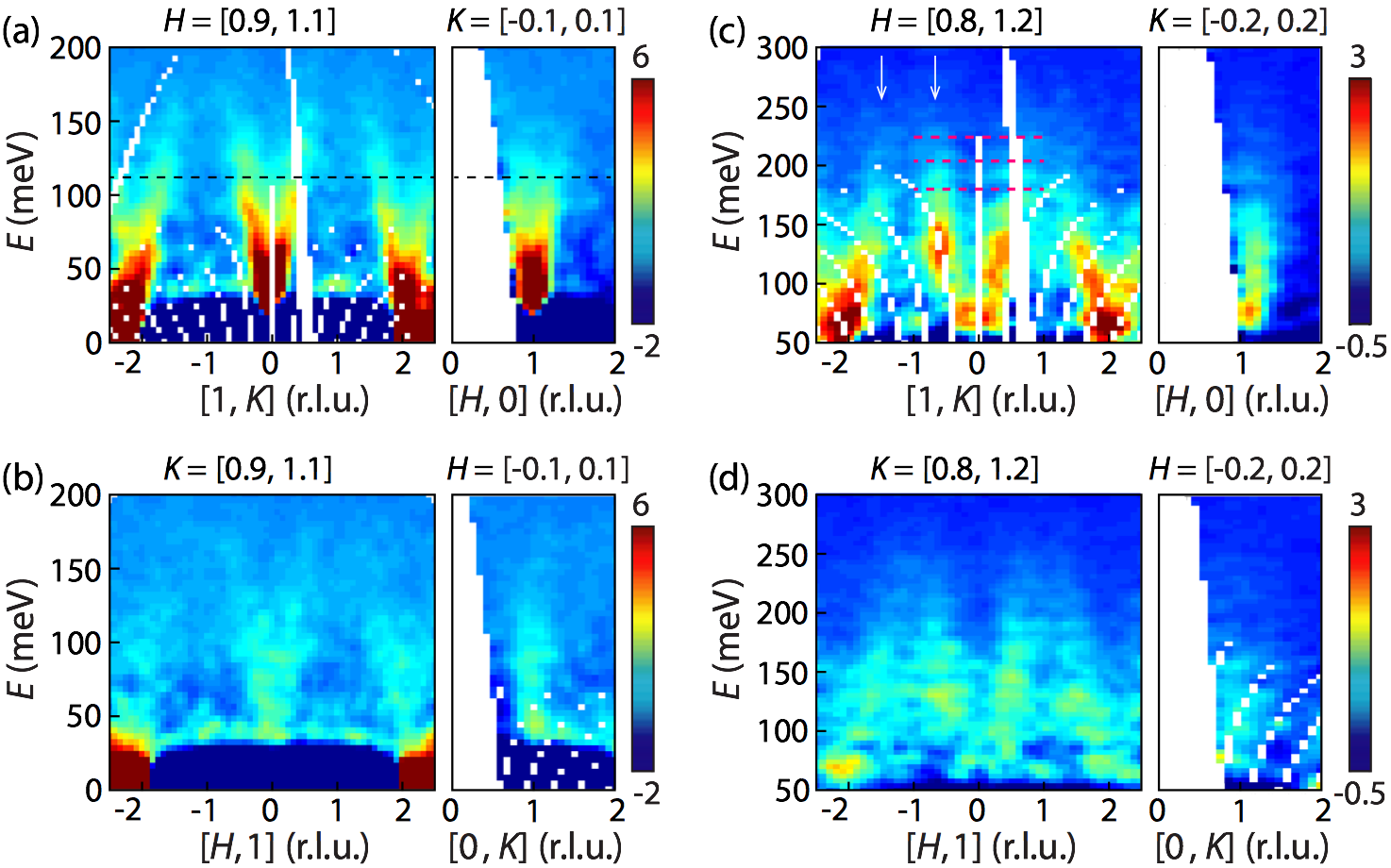}
\caption{Energy-momentum slices of the spin waves in detwinned {\BFA}. Spin waves across $\mathbf{Q}=(1, 0)$ along $[1, K]$ and $[H, 0]$ directions (a), and  $\mathbf{Q}=(0, 1)$ along $[H, 1]$ and $[0, K]$ directions (b), collected with $E_i=250$ meV. The black dashed line in (a) marks an energy above which the spin waves are heavily damped along $[H, 0]$. (c), (d) Similar slices collected with $E_i=450$ meV. The white vertical arrows mark the magnetic excitations persistent at high energy ($E>200$ meV). The pink dashed lines in (c) indicate the energy-momentum trajectories for 1D cuts in Fig. \ref{fig.F3b}.}
\label{fig.F3a}
\end{figure*}

\begin{figure*}[htbp]
\includegraphics[width=12cm]{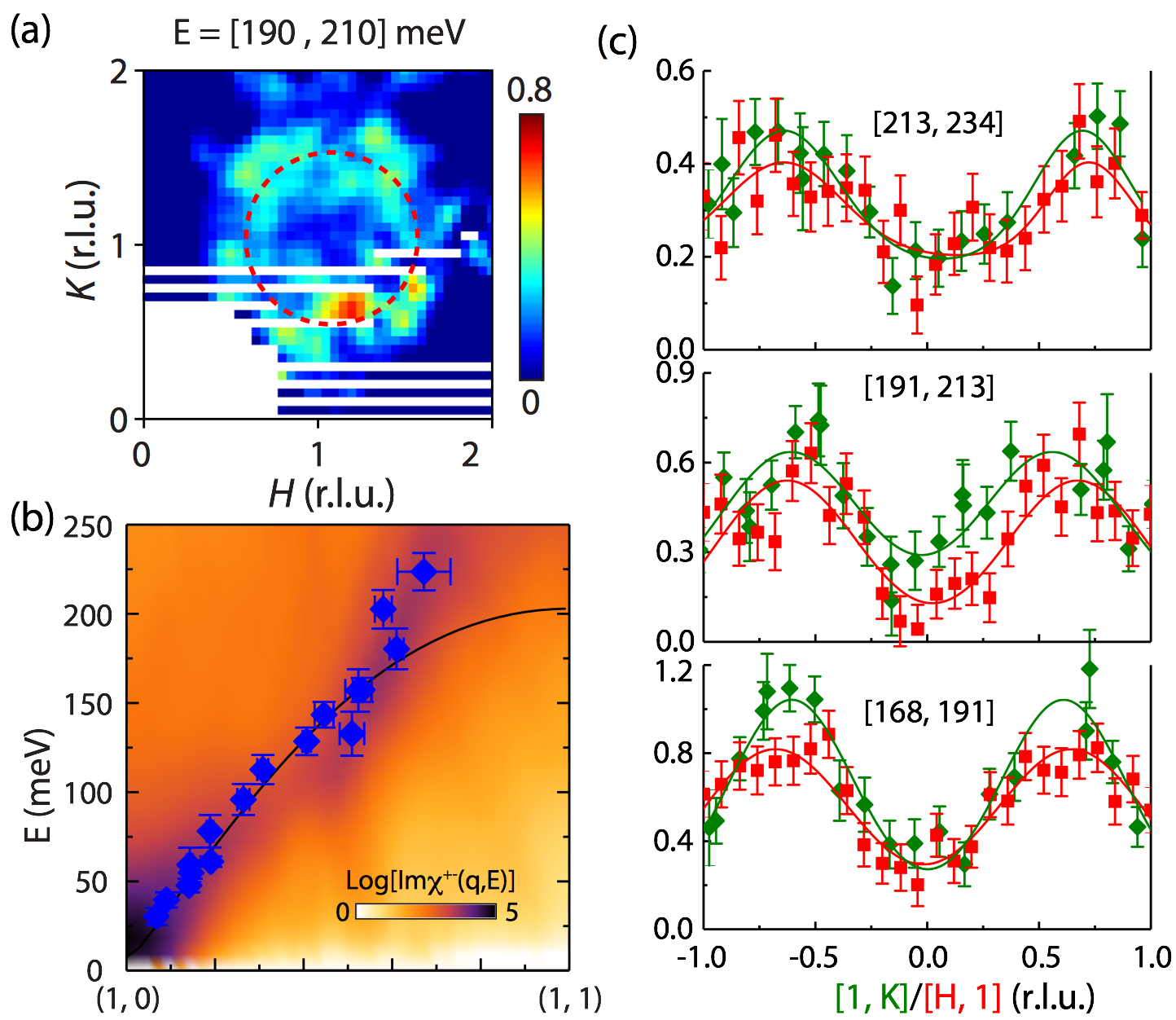}
\caption{Magnon dispersion close to $(1, 1)$. (a) Constant-energy slice with energy transfer $E = 200\pm10$ meV, collected with $E_i=450$ meV. (b) Spin wave dispersion along $[1, K]$ obtained from constant-energy cut. The red and white dashed lines in (a) and (b) are guides to the eye. (c) Constant-energy cuts along $[1, K]$ for $E=[168, 191], [191, 213]$ and $[213, 234]$ meV, which are marked as pink dashed lines in Fig. \ref{fig.F3a}(c).}
\label{fig.F3b}
\end{figure*}

Figures \ref{fig.F3}, \ref{fig.F3a} and \ref{fig.F3b} are detailed one-dimensional (1D) constant-energy cuts and two dimensional slices, from which the damping of spin waves and the magnon dispersion were extracted. 

Fig. \ref{fig.F3} are 1D constant-energy cuts along the transverse and the longitudinal directions across the $(1, 0)$ and $(0, 1)$ positions, corresponding to the constant-energy slices shown in Fig. 2 (main text). The magnetic excitations arise from $(1, 0)$ and disperse out along the transverse and longitudinal directions, forming an elliptical ring in constant energy slices [Fig. 2 and the color map in Fig. \ref{fig.F3}], as also indicated by the twin-peak structure of the cuts [Fig. \ref{fig.F3}]. The magnetic excitation difference between $(1, 0)$ and $(0, 1)$ decrease with increasing energy, in agreement with the constant-energy slices and energy-dependent local susceptibility in Figs. 1 and 2. 

The 1D cuts along transverse directions ($[1, K]$ and $[H, 1]$) show two peaks at all energies (with decreasing correlation length), while the longitudinal cuts show twin peaks only below $\sim 110$ meV [Fig. \ref{fig.F3}], above which the magnetic excitations are heavily damped and form one broad peak with decreasing intensity, which cannot be fitted by two magnetic excitation peaks (not shown here). This is also clearly shown in the dispersion and spectral weight distribution of the spin waves collected with $E_i=250$ and $450$ meV [Figure \ref{fig.F3a}]. The two dispersive branches along transverse direction ($[1, K]$ in Figs. \ref{fig.F3a}(a) and (c)) persist to very high energy ($>200$ meV). However, the spin waves along $[H, 0]$ direction were damped into a weak, broad continuum above $\sim 110$ meV [Fig. \ref{fig.F3a}(a)]. This damping along longitudinal direction has been attributed to the interactions between spin waves and particle-hole excitations in a Stoner continuum. This anisotropic damping along transverse and longitudinal directions has been reported in previous neutron scattering results on {\BFA} \cite{leland11} and NaFeAs \cite{chenglin14prl}. The energy-momentum slices in Fig. \ref{fig.F3a} also show the general damping of the magnetic excitations, that is, broadening of the energy width (decrease of lifetime) of the magnon. This could be driven by interactions between magnetic excitations and other elementary excitations.

Now we turn to the magnon spectral weight and dispersion at high energy around $(1, 1)$, which exhibits several interesting features that have not been well described in the frame of local moment picture.  According to the $J_{1a}-J_{1b}-J_1$ Heisenberg model described in the main text, spin waves along transverse directions ($[H, \pm1]$ and $[\pm1, K]$) converge on their magnon band top ($\sim200$ meV) at $(\pm1, \pm1)$. The local susceptibility $\chi''(E)$ also drop drastically above the band top. 

However, as we have shown in Fig. 2(l) and Fig. \ref{fig.F3}(n), magnetic excitations arising from (1, 0) with $E = 224\pm10$ meV reach only $(1, \pm0.7)$, distinct from the prediction drawn by Heisenberg model. This is even more clear in Fig. \ref{fig.F3a} (c) and (d). Though the spin waves disperse out from $(1, 0)$ and exhibit a dispersion consistent with that from Heisenberg model below $\sim180$ meV, the magnetic excitations above $\sim200$ become almost non-dispersive and persist to very high energy ($\sim300$ meV), giving rise to substantial spectral weight above the ``band top'' ($\sim 200$ meV). 

Momentum cuts is usually not as effective as energy spectra in determining magnon dispersion especially for flat band top. The magnon dispersion shown in Fig. 1(c) is limited to $E<180$ meV to avoid determining the dispersion using momentum cuts close to band top (at (1, 1) with $E = 200$ meV) predicted by Heisenberg model. However, because no flat band top of magnon dispersion was observed in Fig. \ref{fig.F3a}, we tentatively expand the dispersion to higher energy using 1D constant-energy $[1, K]$ cuts. Figure \ref{fig.F3b} summarize the determination of the magnon dispersion in the range of [191, 234] meV using $[1, K]$ cuts. A ring-like signal instead of spot is clearly observed at $200\pm10$ meV, as shown in Fig. \ref{fig.F3b}(a), indicating the prediction of a $\sim 200$meV band top at (1, 1) using Heisenberg model is incorrect. It is further evidenced by the $[1, K]$ cuts in Fig. \ref{fig.F3b}(c) for energy ranges [168, 191], [191, 213] and [213, 234] meV. These 1D cuts provide two more data points in the dispersion [Fig. \ref{fig.F3b}(b)], leading to an upturn on the dispersion. Surprisingly, the new dispersion including the upturn follows the spectral weight color map from RPA calculation [Figs. 1(c), S2 and \ref{fig.F3b}(b)]. The high energy spectral weight above 200 meV can also be qualitatively captured in the same calculation [Fig. 1(d)]. These two features and their comparison with Heisenberg model and RPA calculation suggest that the magnetic excitations are associated with itinerant magnetism instead of local magnetic moments. Therefore, the RPA calculation including Hund's coupling and on-site electron correlation in this manuscript has captured most features of the intrinsic spin waves measured in detwinned {\BFA}.

\subsection{Nematic spin correlations in the tetragonal state of uniaxial-strained {\BFA}}

\begin{figure}[htbp]
\includegraphics[width=14.5cm]{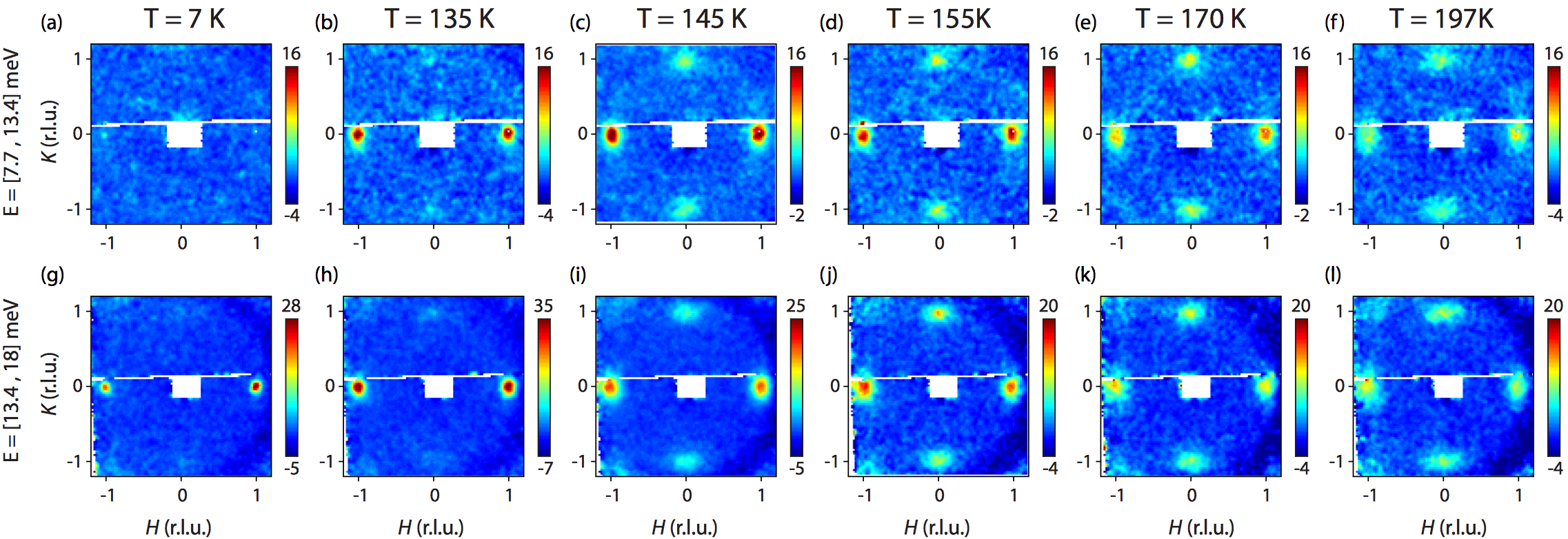}
\caption{ Constant energy slices of the magnetic excitations for {\BFA} under $P \sim 12-22$ MPa uniaxial pressures measured at $T=7$K, $135$K, $145$K, $155$K, $170$K and $197$K, with $E_i=30$ meV.}
\label{fig.F4a}
\end{figure}

The dynamic susceptibility difference between $\mathbf{Q_1}=(1, 0)$ and $\mathbf{Q_2}=(0, 1)$ in the paramagnetic state of uniaxial-strained {\BFA}, termed nematic spin correlations $\psi(E, T)=[\chi''_1(E)-\chi''_2(E)]/[\chi''_1(E)+\chi''_2(E)]$, has been discussed in the main text. In this section, we present more detailed data complementary to the Fig. 4 of the main text. 

\begin{figure*}[htbp]
\includegraphics[width=14cm]{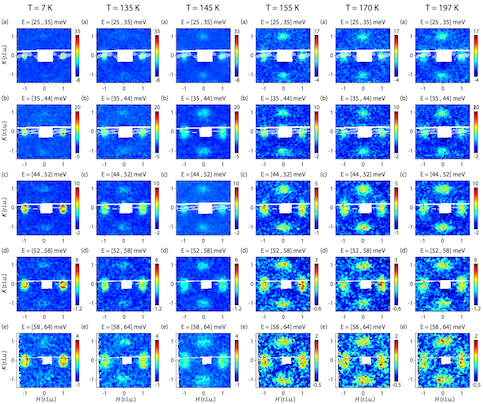}
\caption{Constant energy slices of the magnetic excitations for {\BFA} under $P \sim 12-22$ MPa uniaxial pressures measured at $T=7$K, $135$K, $145$K, $155$K, $170$K and $197$K, with $E_i=81$ meV.}
\label{fig.F4b}
\end{figure*}

\begin{figure*}[htbp]
\includegraphics[width=14cm]{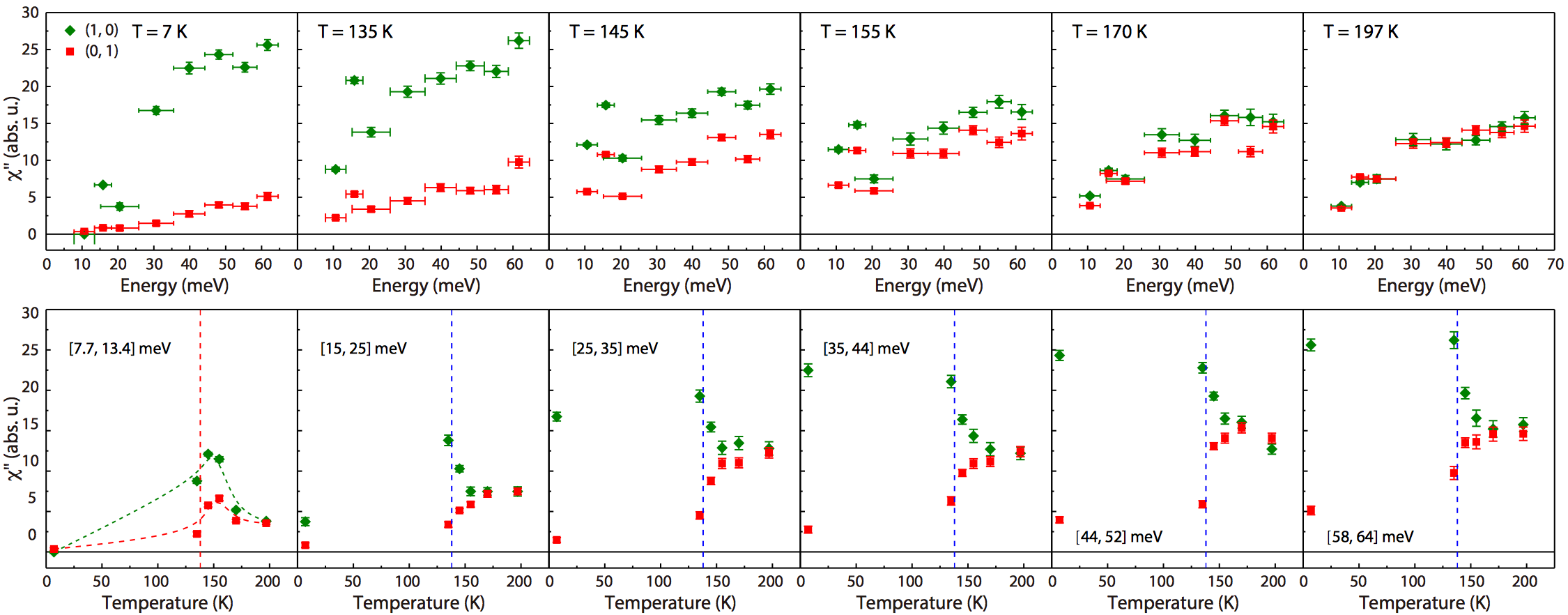}
\caption{Energy and temperature dependence of the local susceptibility below $64$ meV, extracted from the slices shown in Figs. \ref{fig.F4a} and \ref{fig.F4b} measured with $E_i=30$ and $81$ meV. }
\label{fig.F5}
\end{figure*}

The uniaxial pressures applied on the 16 pieces of {\BFA} crystals are roughly estimated to be $~12-22$ MPa, which could vary slightly with temperature because of the thermal contraction of the device. Since the significant effect of uniaxial pressure occurs within a narrow temperature range close to $T_S$, in which we assume the pressure would not change much. Because the pressures were difficult to be precisely controlled using spring washers, the effect of pressure here is an average of that from all the samples.   The effect of a $\sim20$MPa uniaxial pressure on the structure of a single piece of {\BFA} has been shown in Fig. \ref{fig.F1}(b). Magnetic transition temperature could be enhanced to about 143K under $P=20$ MPa \cite{haoran15, david17}. Figs. \ref{fig.F4a} and \ref{fig.F4b} show the temperature dependent magnetic excitations below $18$ meV measured with $E_i=30$ meV and high energy transfers measured with $E_i=81$ meV, respectively. Fig. \ref{fig.F5} summarizes the temperature and energy dependence of the local susceptibility extracted from Figs. \ref{fig.F4a} and Fig. \ref{fig.F4b}, which illustrates the energy and temperature evolution of the nematic spin correlations clearly.  

\begin{figure}[htbp]
\includegraphics[width=10cm]{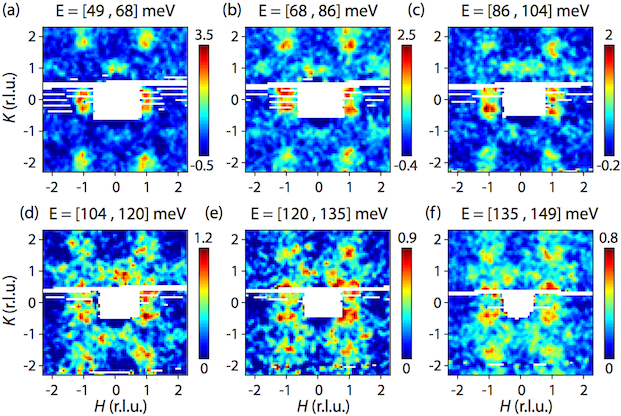}
\caption{Spin waves measured at $T=135$K with $E_i=250$ meV.}
\label{fig.F6}
\end{figure}
\begin{figure*}[htbp]
\includegraphics[width=14cm]{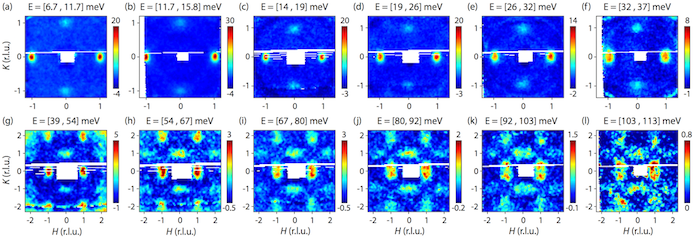}
\caption{Constant energy slices of the magnetic excitations measured at $T=140$K with $E_i=25$ (a-b), $50$ (c-f) and $163$ (g-l) meV.}
\label{fig.F7}
\end{figure*}
\begin{figure}[htbp]
\includegraphics[width=10cm]{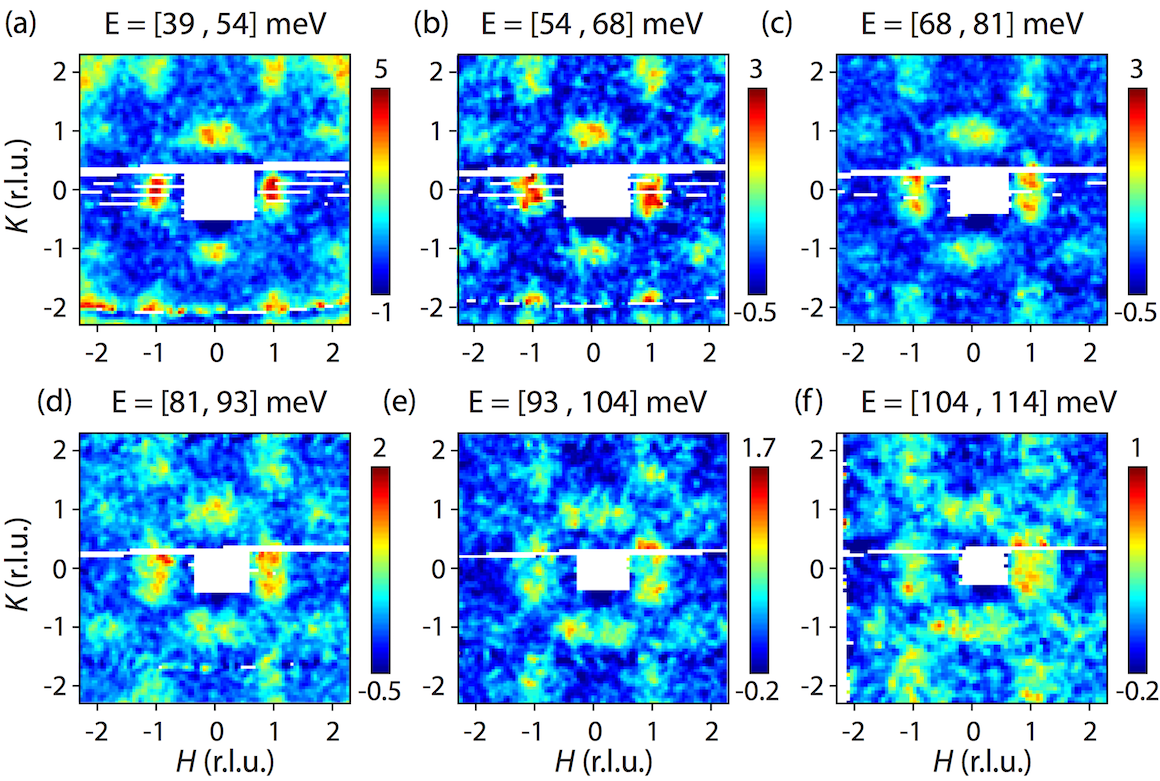}
\caption{ Constant energy slices of the magnetic excitations measured at $T=145$K with $E_i=165$meV.}
\label{fig.F8}
\end{figure}

The spin excitation anisotropy $\psi(E)$ persist in high energy transfers at temperatures below and slightly above $T_N$. Therefore, we have carried out neutron scattering measurements using $E_i=250$ meV for $T=135$K (Fig. \ref{fig.F8}), $E_i=163$ meV for $T=140$ K (multi-$E_i$ chopper with sub-$E_i = 50$ and $25$ meV), and $E_i=165$ meV for $T=145$K, to determine the energy scales of $\psi(E)$. The results extracted from Figs. \ref{fig.F4a} to \ref{fig.F8} have been summarized in Fig. 4 of the main text.

\newpage

%-------------------------------------------------------------------------------------

%-------------------------------------------------------------------------------------

\end{document}